\documentclass[aps,prb,twocolumn]{revtex4}
\usepackage{graphics}
\usepackage{graphicx}
\begin{document}
\title{An Overview of Multiscale Simulations of Materials}
\author{Gang Lu and Efthimios Kaxiras}
\affiliation{
Department of Physics and Division of Engineering and Applied Science,\\
 Harvard University, Cambridge, MA 02138}
\begin{abstract}
Multiscale modeling of material properties has emerged as one of
the grand challenges in material science and engineering. We
provide a comprehensive, though not exhaustive, overview of the
current status of multiscale simulations of materials.  We
categorize the different approaches in the spatial regime into
{\em sequential} and {\em concurrent}, and we discuss in some
detail representative methods in each category. We classify the
multiscale modeling approaches that deal with the temporal scale
into three different categories, and we discuss representative
methods pertaining to the each of these categories. Finally, we
offer some views on the strength and weakness of the multiscale
approaches that are discussed, and touch upon some of the
challenging multiscale modeling problems that need to be addressed
in the future.

\end{abstract}
\maketitle
\section{Introduction}

Some of the most fascinating problems in all fields of science
involve multiple spatial and/or temporal scales: processes that
occur at a certain scale govern the behavior of the system across
several (usually larger) scales.  The notion and practice of
multiscale modeling can be traced back to the beginning of modern
science (see, for example, the discussion in \cite{rob}). In many
problems of materials science this notion arises quite naturally:
the ultimate microscopic constituents of materials are atoms, and
the interactions among them at the microscopic level (of order
nanometers and femtoseconds) determine the behavior of the
material at the macroscopic scale (of order centimeters and
milliseconds and beyond), the latter being the scale of interest
for technological applications. The idea of performing simulations
of materials across several characteristic length and time scales
has therefore obvious appeal as a tool of potentially great impact
on technological innovation \cite{kaxiras,mrs,yip}. The advent of
ever more powerful computers which can handle such simulations
provides further argument that such an approach can address
realistic situations and can be a worthy partner to the
traditional approaches of theory and experiment.

In the context of materials simulations, one can distinguish four
characteristic length levels:\\
(1) The {\em atomic} scale ($\sim 10^{-9}$m or a few nanometers)
in which the electrons are the players, and their
quantum-mechanical state dictates the interactions among the atoms.\\
(2) The {\em microscopic} scale ($\sim 10^{-6}$m or a few
micrometers) where atoms are the players and their interactions
can be described by classical interatomic potentials (CIP) which
encapsulate the effects of bonding between them, which is mediated by electrons.\\
(3) The {\em mesoscopic} scale ($\sim 10^{-4}$m or hundreds of
micrometers) where lattice defects such as dislocations, grain
boundaries, and other microstructural elements are the players.
Their interactions are usually derived from phenomenological
theories which encompass the effects of interactions between the atoms.\\
(4)The {\em macroscopic} scale ($\sim 10^{-2}$m or centimeters and
beyond) where a constitutive law governs the behavior of the
physical system, which is viewed as a continuous medium. 
In the macroscale, continuum fields such as density, velocity,
temperature, displacement and stress fields, etc
are the players. 
The constitutive laws are usually formulated so that 
they can capture the effects on materials properties from
lattice defects and microstructural elements.\\
Phenomena at each length scale typically have a corresponding time
scale which, in correspondence to the four length-scales mentioned
above, ranges roughly from femtoseconds to picoseconds, to
nanoseconds to milliseconds and beyond.

At each length and time-scale, well-established and efficient
computational approaches have been developed over the years to
handle the relevant phenomena. To treat electrons explicitly and
accurately at the atomic scale, methods known as Quantum Monte
Carlo (QMC)\cite{qmc} and Quantum Chemistry (QC)\cite{qc} can be
employed, which are computationally too demanding to handle more
than a few tens of electrons. Methods based on density functional
theory (DFT) and local density approximation (LDA) \cite{dft,payne} 
in its various implementations, while
less accurate than QMC or QC methods, can be readily applied to
systems containing several hundred atoms for static properties.
Dynamical simulations with DFT methods are usually limited to
time-scales of a few picoseconds. For materials properties that
can be modeled reasonably well with a small number of atoms (such
as bulk crystal properties or point defects), the DFT approach can
provide sufficiently accurate results. Recent progress in linear
scaling electronic structure methods \cite{linear} has enabled
DFT-based calculations to deal with a few thousands atoms
(corresponding to sizes of a couple of nanometers on a side) with
adequate accuracy. Finally, the semi-classical tight-binding
approximation (TBA), although typically not as accurate as DFT
methods, can extend the reach of simulations to a few nanometers
in linear size and a few nanoseconds in time-scale for the
dynamics \cite{colombo}.

For material properties at the microscopic scale, Molecular
Dynamics (MD) and Monte Carlo (MC) simulations are usually
performed employing CIP which can often be derived from DFT
calculations \cite{bazant,eam}. Although not as accurate as the
DFT and TBA methods, the classical simulations are able to provide
insight into atomic processes involving considerably larger
systems, reaching up to $\sim 10^{9}$ atoms \cite{seager}. The
time-scale that MD simulations based on CIP can reach is up to a
microsecond.

At the mesoscopic scale, the atomic degrees of freedom are not
explicitly treated, and only larger scale entities are modeled.
For example, in what concerns the mechanical behavior of solids,
dislocations are the objects of interest. In treating
dislocations, recent progress has been concentrated on the
so-called Dislocation Dynamics (DD) approach
\cite{kubin,hirth2,schwarz,needleman} which has come to be
regarded as one of the most important developments in
computational materials science and engineering in the past two
decades \cite{vasily}. Such DD models deal with the kinetics of
dislocations and can study systems with a few tens of microns in
size and with a maximum strain $\sim$ 0.5\% for a strain rate of
10 sec$^{-1}$ in bcc metals \cite{rhee}.

Finally, for the macroscopic scale, finite-element (FE) methods
\cite{fe} are routinely used to examine the large-scale properties
of materials considered as an elastic continuum \cite{dawson}. For
example, FE methods have been brought to bear on problems of
strain-gradient plasticity, such as geometrically necessary
dislocations \cite{ohashi}. Continuum Navier-Stoke equations are
also often used to study fluids. 

The challenge in modern simulations of materials science and
engineering is that real materials usually exhibit phenomena at
one scale that require a very accurate and computationally
expensive description, and phenomena at another scale for which a
coarser description is satisfactory and in fact necessary to avoid
prohibitively large computations. Since none of the methods above
alone would suffice to describe the entire system, the goal
becomes to develop models that combine different methods
specialized at different scales, effectively distributing the
computational power where it is needed most. It is the hope that
a multiscale approach is
the answer to such a quest, and it is by definition an approach
that takes advantage of the multiple scales present in a material
and builds a unified description by linking the models at the
different scales. Fig. 1 illustrates the concept of a unified
multiscale approach which can reach the length and time scale that
individual methods, developed to treat a particular scale
accurately, fail to achieve. At the same time, the unified
approach can retain the accuracy that the individual approaches
provide in their respective scales, allowing, for instance, for
very high accuracy in particular regions of the systems as
required. As effective theories, multiscale models are also useful
for gaining physical insight that might not be apparent from brute
force computations. Specifically, a multiscale model can be an
effective way to facilitate the reduction and the analysis of data
which sometimes can be overwhelming. Overall, the goal of
multiscale approaches is to predict the performance and behavior
of materials across all relevant length and time scales, striving
to achieve a balance between accuracy, efficiency and realistic
description.

Conceptually, two categories of multiscale simulations can be
envisioned, sequential and concurrent. The sequential methodology
attempts to piece together a hierarchy of computational approaches
in which large-scale models use the coarse-grained representations
with information obtained from more detailed, smaller-scale
models. This sequential modeling approach has proven effective in
systems where the different scales are weakly coupled. The
characteristic of the systems that are suited for a sequential
approach is that the large-scale variations decouple from the
small-scale physics, or the large-scale variations appear
homogeneous and quasi-static from the small-scale point of view.
Sequential approaches have also been referred to as serial,
implicit or message-passing methods. The vast majority of
multiscale simulations that are actually in use are sequential.
Examples of such approaches abound in literature, including almost
all MD simulations whose underlying potentials are derived from
electronic structure theory \cite{pettifor1,pettifor2}, usually
{\it ab initio} calculations \cite{bazant,eam}. One frequently
mentioned \cite{abraham1, abraham2} example of sequential
multiscale simulations is the work of Clementi {\it et al.}
\cite{clementi} who used QC calculations to evaluate the
interaction of several water molecules; from this database, an
empirical potential was parameterized for use in molecular
dynamics simulations; the MD simulations were then used to
evaluate the viscosity of water from atomic autocorrelation
functions; and finally, the computed viscosity was employed in
computational fluid dynamics calculations to predict the tidal
circulation in Buzzard's Bay of Massachusetts.

The second category of multiscale simulations consists of the
so-called concurrent, or parallel, or explicit approaches. These
approaches attempt to link methods appropriate at each scale
together in a combined model where the different scales of the
system are considered concurrently and communicate with some type
of hand-shaking procedure. This approach is necessary for systems
that are inherently multiscale, that is, systems whose behavior at
each scale depends strongly on what happens at the other scales.
In contrast to sequential approaches, the concurrent simulations
are still relatively new and only a few models have been developed
to date. In a concurrent simulation, the system is often
partitioned into domains characterized by different scales and
physics. The challenge of the concurrent approach lies at the
coupling between the different regions treated by different
methods, and a successful multiscale model seeks a smooth coupling
between these regions.

In principle, multiscale simulations could be based on a hybrid
scheme, using elements from both the sequential and the concurrent
approaches.  We will not examine this type of approach in any
detail, since it involves no new concepts other than the
successful combination of elements underlying the other two types
of approaches.

There already exist a few review papers and special editions of
articles on multiscale simulation of materials in literature
\cite{kaxiras,mrs,bullard,ghoniem,robertson,kubin2,campbell}. 
A mathematic perspective of multiscale modeling and computation  
is also available \cite{weinan_ams}. The
present overview does not aim to provide another collection of
various multiscale techniques, but rather to identify the
characteristic features and classify multiscale simulation
approaches into rational categories in relation to the problems
where they apply. We select a few illustrative examples for each
category and try to establish connections between these approaches
whenever possible. Since almost all interesting material behavior
and processes are time dependent, we will address both the issue
of length-scales and the issue of time-scales integration in
materials modeling.

The paper is organized as follows: In Section II we discuss in
detail representative examples of sequential multiscale approaches
in the spatial regime.  In Section III we present examples of
concurrent multiscale approaches, also in the spatial regime . In
Section IV we discuss representative approaches that extend
time-scales in dynamical simulations. Section V contains our
comments and conclusions for the applicability of the various
approaches. The examples presented in this overview to some extent
reflect our own research interests and they are by no means
exhaustive. Nevertheless, we hope that they give a satisfactory
cross-section of the current state of the field, and they can
serve as inspiration for further developments in this exciting
endeavor.

\section{Sequential multiscale approaches}

Two ingredients are required in order to construct a successful
sequential multiscale model:\\
(i) It is necessary to have {\it a priori} and complete knowledge
of the fundamental processes at the lowest scale involved. This
knowledge or information can then be used for modeling the system
at successively coarser scales.\\
(ii) It is necessary to have a reliable strategy for encompassing
the lower-scale information into the coarser scales.  This is
often accomplished by phenomenological theories, which contain a
few key parameters (these can be functions), the value of which is
determined from the information at the lower scale.\\
This message-passing approach can be performed in sequence for
multiple length scales, as in the example cited in the
introduction\cite{clementi}. The key attribute of the sequential
approach is that the simulation at a higher level critically
depends on the completeness and the correctness of the information
gathered at the lower level, as well as the efficiency and
reliability of the model at the coarser level.

To illustrate this type of approach, we will present two examples
of sequential multiscale approaches in some detail. The first
example concerns the modeling of dislocation properties in the
context of the Peierls-Nabarro (P-N) phenomenological model, where
the lower scale information is in the form of the so-called
generalized stacking fault energy surface (also referred to as the
$\gamma$-surface), and the coarse-grained model is a
phenomenological continuum description. The second example
concerns the modeling of coherent phase transformations in the
context of the phase-field approach, where the lower scale
knowledge is in the form of {\it ab initio} free energies, and the
coarse-grained model is again a continuum model.

\subsection{Peierls-Nabarro model of dislocations}

Dislocations are central to our understanding of mechanical
properties of crystalline solids. In particular, the creation and
motion of dislocations mediate the plastic response of a crystal
to external stress. While continuum elasticity theory describes
well the long-range elastic strain of a dislocation for length
scales beyond a few lattice spacings, it breaks down in the
immediate vicinity of the dislocation core. There has been a great
deal of interest in describing accurately the dislocation core
structure on an atomic scale because the core structure to a large
extent dictates the dislocation properties \cite{richardson,
vitek}. So far, direct atomistic simulation of dislocation
properties based on CIP has not been satisfactory because the CIP
is not always reliable or may even not be available for the
material of interest, especially when the physical system involves
several types of atoms. On the other hand, {\it ab initio}
calculations are still computationally expensive for the study of
dislocation core properties, particularly of dislocation mobility.
Recently, a promising approach based on the framework of the
Peierls-Nabarro (P-N) model has attracted considerable interest
for the study of dislocation core structure and mobility
\cite{joos1,joos2,juan,hartford,sydow,medvedeva,
mryasov,bulatov1,lu1,lu2,lu3,lu4}. This approach when combined
with {\it ab initio} calculations for the energetics, represents a
plausible alternative to the direct {\it ab initio} simulations of
dislocation properties.

The P-N model is an inherently multiscale framework, first
proposed by Peierls \cite{peierls} and Nabarro \cite{nabarro} to
incorporate the details of a discrete dislocation core into an
essentially continuum framework. Consider a solid with an edge
dislocation in the middle (Fig. \ref{edge}): the solid containing
this dislocation is represented by two elastic half-spaces joined
by atomic level forces across their common interface, known as the
glide plane (dashed line). The goal of the P-N model is to
determine the slip distribution 
on the glide plane, which minimizes the total energy.
The dislocation is characterized by the slip (relative
displacement) distribution
\begin{equation}
{\bf f}(x) = {\bf u}(x,0^{+}) - {\bf u}(x,0^{-})
\label{misfit}
\end{equation}
which is a measure of the misfit across the glide plane; ${\bf
u}(x,0^+)$ and ${\bf u}(x,0^{-})$ are the displacement of the
half-spaces at position $x$ immediately above and below the glide
plane. The total energy of the dislocated solid includes two
contributions: (1) the nonlinear potential energy due to the
atomistic interaction across the glide plane, and (2) the elastic
energy stored in the two half-spaces associated with the presence
of the dislocation. Both energies are functionals of the slip
distribution ${\bf f}(x)$. Specifically, the nonlinear misfit
energy can be written as
\begin{equation}
U_{misfit} = \int^{\infty}_{-\infty}\gamma({\bf f}(x))dx,
\label{Umisfit}
\end{equation}
where $\gamma({\bf f})$ is the generalized stacking fault energy
surface (the $\gamma$-surface) introduced by Vitek \cite{vitek1}.
The nonlinear interplanar $\gamma$-surface can, in general, be
determined from atomistic calculations. For systems where CIP are
not available or not reliable (for instance, it is exceedingly
difficult to derive reliable potentials for multi-component
alloys), {\it ab initio} calculations can be performed to obtain
the $\gamma$-surface. On the other hand, the elastic energy of the
dislocation can be calculated reasonably from elasticity theory:
the dislocation may be thought of as a continuous distribution of
infinitesimal dislocations whose Burgers vectors integrate to that
of the original dislocation \cite{eshelby0}. Therefore, the
elastic energy of the original dislocation is just the sum of the
elastic energy due to all the infinitesimal dislocations (from the
superposition principle of linear elasticity theory), which can be
written as
\begin{equation}
U_{elastic} = \frac{\mu}{2\pi(1-\nu)}\int dx \int dx'
\ln\frac{L}{|x-x'|} \frac{d{\bf f}(x)}{dx}\frac{d{\bf f}(x')}{dx'}
\label{Uelast}
\end{equation}
where $\mu$, and $\nu$ are the shear modulus and Poisson's ratio,
respectively. $L$ is an inconsequential constant introduced as a
large-distance cutoff for the computation of the logarithmic
interaction energy \cite{hirth}. Note that the Burgers vector of
each infinitesimal dislocation is the local gradient in the slip
distribution, $d{\bf f}(x)/dx$. The gradient of ${\bf f}(x)$ is
called dislocation (misfit) density, denoted by $\rho(x)$. Since
the P-N model requires that atomistic information (embodied in the
$\gamma$-surface) be incorporated into a coarse-grained continuum
framework, it is a sequential multiscale strategy. Thus the
successful application of the method depends on the reliability of
both $\gamma$-surface and the underlying elasticity theory which
is the basis for the formulation of the phenomenological theory.

In the current formulation, the total energy is a functional of
misfit distribution ${\bf f}(x)$ or equivalently $\rho(x)$, and it
is invariant with respect to arbitrary translation of $\rho(x)$
and ${\bf f}(x)$. In order to regain the lattice discreteness, the
integration of the $\gamma$-energy in Eq. (\ref{Umisfit}) was
discretized and replaced by a lattice sum in the original P-N
formulation,
\begin{equation}
U_{misfit} = \sum_{i=-\infty}^{\infty} \gamma({\bf f}_i)\Delta x,
\label{discrete}
\end{equation}
with $x_i$ the reference position and $\Delta x$ the average
spacing of the atomic rows in the lattice. This procedure,
however, is inconsistent with evaluation of elastic energy [Eq.
(\ref{Uelast})] as a continuous integral. Therefore the total
energy is {\it not} variational. Furthermore in the original P-N
model, the shape of the solution ${\bf f}(x)$ is assumed to be
invariant during dislocation translation, a problem that is also
associated with the non-variational formulation of the total
energy.

To resolve these problems, a so-called Semidiscrete Variational
P-N (SVPN) model was recently developed\cite{bulatov1}, which
allows for the first time the study of narrow dislocations, a
situation that the standard P-N model can not handle. Within this
approach, the equilibrium structure of a dislocation is obtained
by minimizing the dislocation energy functional
\begin{equation}
U_{disl} = U_{elastic} + U_{misfit} + U_{stress} + Kb^2lnL,
\label{pn}
\end{equation}
where
\begin{equation}
U_{elastic} = \sum_{i,j}\frac{1}{2}\chi_{ij}[K_e(\rho_i^{(1)}\rho_j^{(1)} +
\rho_i^{(2)}\rho_j^{(2)}) + K_s\rho_i^{(3)}\rho_j^{(3)}],
\end{equation}
\begin{equation}
\label{gamma}
U_{misfit} =  \sum_i\Delta x \gamma_3({\bf f}_i),
\end{equation}
\begin{equation}
U_{stress} = - \sum_{i,l}\frac{x_i^2-x_{i-1}^2}{2}(\rho_i^{(l)}\tau_i^{(l)}),
\end{equation}
with respect to the dislocation misfit density. Here,
$\rho_i^{(1)}$, $\rho_i^{(2)}$ and $\rho_i^{(3)}$ are the edge,
vertical and screw components of the general interplanar misfit
density at the $i$-th nodal point, and $\gamma_3({\bf f}_i)$ is
the corresponding three-dimensional $\gamma$-surface. The
components of the applied stress interacting with the
$\rho_i^{(1)}$, $\rho_i^{(2)}$ and $\rho_i^{(3)}$, are $\tau^{(1)}
= \sigma_{21}$, $\tau^{(2)} = \sigma_{22}$ and $\tau^{(3)} =
\sigma_{23}$, respectively. $K$, $K_e$ and $K_s$ are the
pre-logarithmic elastic energy factors \cite{hirth}. The
dislocation density at the $i$-th nodal point is $\rho_i = (f_i -
f_{i-1})/(x_i - x_{i-1})$ and $\chi_{ij}$ is the elastic energy
kernel \cite{bulatov1}.

The first term in the energy functional, $U_{elastic}$ is now
discretized in order to be consistent with the discretized misfit
energy, which makes the total energy functional variational.
Another modification in this approach is that the nonlinear misfit
potential in the energy functional, $U_{misfit}$, is a function of
all three components of the nodal misfit, ${\bf f}(x_i)$. Namely,
in addition to the misfit along the Burgers vector, lateral and
even vertical misfits across the glide plane are also included.
This allows the treatment of straight dislocations of arbitrary
orientation in arbitrary glide planes. Furthermore, because the
misfit vector ${\bf f}(x_i)$ is allowed to change during the
process of dislocation translation, the energy barrier (referred
to as the Peierls barrier) can be significantly lowered compared
to its corresponding value from a rigid translation. The response
of a dislocation to an applied stress is achieved by minimization
of the energy functional with respect to $\rho_i$ at the given
value of the applied stress, $ \tau_i^{(l)}$. An instability is
reached when an optimal solution for $\rho_i$ no longer exists,
which is manifested numerically by the failure of the minimization
procedure to converge. The Peierls stress is defined as the
critical value of the applied stress which gives rise to this
instability.

The SVPN model has been applied to study various interesting
material problems related to dislocation phenomena
\cite{lu2,lu3,lu4}. One study involved the understanding of
hydrogen-enhanced local plasticity (HELP) in Al. HELP is regarded
as one of three general mechanisms responsible for H embrittlement
of metals \cite{help}. There has been overwhelming experimental
evidence in support of HELP, but a theoretical foundation was
lacking. In order to gain understanding of the physics behind the
HELP mechanism, Lu {\it et al.} carried out {\it ab initio}
calculations for the $\gamma$-surface of Al with H impurities
placed at the interstitial sites \cite{lu2}. The $\gamma$-surface
for both pure Al and the Al+H systems is shown in Fig.
\ref{gamma2}. Comparing the two $\gamma$-surfaces, one finds an
overall reduction in $\gamma$ energy in the presence of H, which
is attributed to the change of atomic bonding across the glide
plane, from covalent-like to ionic-like \cite{daniel}.

The core properties of four different dislocations, screw
(0$^\circ$), 30$^\circ$, 60$^\circ$ and edge (90$^\circ$) have
been studied using the SVPN model combined with the {\it ab
initio} determined $\gamma$-surface. It was found that the Peierls
stress for these dislocations is reduced by more than an order of
magnitude in the presence of H \cite{lu2}, which is compatible
with the experimental findings that support the HELP
mechanism\cite{help}. Moreover, in order to address the
experimental observation for H trapping at dislocation cores and
H-induced slip planarity, the H binding energy to the dislocation
cores was calculated\cite{lu2}.  These calculations showed that
there is strong binding between H and dislocation cores, that is,
H is attracted (trapped) to dislocation cores which lowers the
core energies. More importantly, the binding energy was found to
be a function of dislocation character, with the edge dislocation
having the greatest and the screw dislocation having the lowest
binding energy. For a mixed dislocation, the binding energy
increases with the amount of edge component of the Burgers vector.
These results suggest that in the presence of H, it costs more
energy for an edge dislocation to transform into a screw
dislocation in order to cross-slip, since the edge dislocation has
almost twice the binding energy of the screw dislocation
\cite{lu2}. In the same vein, it costs more energy for a mixed
dislocation to transfer its edge component to a screw component
for cross-slip. Therefore the cross-slip process is suppressed due
to the presence of H, and slip is confined to the primary glide
plane, exhibiting the experimentally observed slip planarity.

A similar approach was applied to the study of the vacancy
lubrication effect on dislocation motion in Al. From this
analysis, it was shown that the role of vacancies is crucial in
reconciling the results of Peierls stress measured from different
experimental techniques \cite{lu3}. Very recently, a multiple
plane P-N model has been developed to study dislocation phenomena
involving more than one glide planes, such as dislocation
constriction and cross-slip \cite{lu4}. Finally we should point
out that the P-N model is just one example of more general
cohesive surface models that are built upon the idea of limiting
all constitutive nonlinearity to certain privileged interfaces,
while the remainder of the material is treated via more
conventional continuum theories \cite{rob}. The same strategy can
also be applied to the study of fracture and dislocation
nucleation from a crack tip \cite{guanshui}.

It is interesting to note that the analysis of $\gamma$-surface
can provide a qualitative understanding of even more complex
mechanical properties of materials. For example, Rice and
coworkers \cite{Rice} formulated powerful criteria for the brittle
behavior of materials, by extending the Peierls analysis to
geometries involving cracks.  Based on this framework, Waghmare
{\it et al.}\cite{umesh1,umesh2} were able to predict which
alloying elements could improve the ductility of MoSi$_2$ by
analyzing the {\it ab initio} determined $\gamma$-surface of the
alloys, and comparing the changes induced by alloying to key
features of the $\gamma$-surface versus the changes induced to the
surface energy $\gamma_s$. Remarkably, certain predictions of this
relatively simple theoretical modeling were borne out by
subsequent experiments \cite{peralta}.

We have devoted some attention to the description of the P-N model
and its implementation using {\it ab initio} $\gamma$-surfaces,
because it is an ideal case of a sequential multiscale model: it
consists of a well motivated phenomenological framework, within
which the set of atomistically derived quantities is well defined
and complete (in this case the $\gamma$-surface).  In this sense,
it fulfills all the requirements for a coherent and complete
multiscale model. There are no doubt limitations to it, arising
from the range of validity of the phenomenological theory, but
within this range there are no other ambiguities in constructing
the multiscale model.  Perhaps, its successes, some of which we
presented above, are owed to this complete character of the model.

\subsection{Phase-field model of coherent phase transformations}

The structure-properties paradigm is one of the principal pillars
in materials science. The term ``structure'' here refers to
structures at many different scales, including the atomic scale
geometry determined by the crystalline arrangement of atoms, the
structure of the defects that exist in a material, and the
structure that emerges as a result of the organization of these
defects into what is referred to as microstructure. Among these
structures, the microstructure on the scale of micrometers is
often directly tied to the mechanical properties of materials, and
has therefore attracted great interest both in terms of scientific
understanding and practical applications \cite{suo,cocks,gurtin,
bullard1,fan}.

Recently, a powerful sequential multiscale approach has been put
forward for modeling the precipitate microstructure and its
evolution in multicomponent alloys \cite{chen1,vaith}, materials
which appear in many technological applications. The approach is
based on the continuum phase-field model whose driving forces
(free energies) are obtained from combined {\it ab initio}
calculations and the mixed-space cluster expansion technique. One
interesting application of this approach concerned the study of
precipitation of the $\theta'$(Al$_2$Cu) phase in Cu-Al alloys
during thermal aging \cite{vaith}.

In the phase-field multiscale approach, the nature of phase
transformation as well as the microstructures that are produced
are described by a set of continuous order-parameter fields
\cite{wang,chen2}. The temporal microstructure evolution is
obtained from solving field kinetic equations that govern the
time-dependence of the spatially inhomogeneous order-parameter
fields. Within the diffuse-interface description, the
thermodynamics of a phase transformation and the accompanying
microstructure evolution are modeled by a free energy that is a
function of the order-parameter field, or phase field. For a
structural transformation, the total free energy can be written
as:
\begin{equation}
{\cal F}_{tot} = {\cal F}_{bulk} + {\cal F}_{inter} + {\cal F}_{elast}
\label{II-B-1}
\end{equation}
where ${\cal F}_{bulk}$ is the bulk free energy, ${\cal F}_{inter}$ is the
interfacial free energy, and ${\cal F}_{elast}$ is the coherency elastic
strain energy arising from the lattice accommodation along the
coherent interfaces in a microstructure. For a microstructure
described by a composition field $c$ and a set of structural
order-parameters, $\eta_1$, ..., $\eta_p$, the first two terms of
Eq. (\ref{II-B-1}) are given by
\begin{eqnarray}
\label{array1}
{\cal F}_{bulk} + {\cal F}_{inter}& =& \int_{V}\{f[c({\bf r}),\eta_p({\bf
 r})]+\frac{\alpha}{2}|\nabla c({\bf r})|^2 \\ \nonumber
& &  + \frac{1}{2}
\sum\limits_{p}\beta_{ij}(p)\nabla_{i}\eta_p({\bf r})
\nabla_{j}\eta_p({\bf r})\}dV
\end{eqnarray}
where $f(c,\eta_p)$ is the local free energy density \cite{li}
and $\alpha$ and $\beta_{ij}(p)$ are the gradient energy coefficients
which control the width of the diffuse interface. The elastic strain
energy is obtained from elasticity theory using the homogeneous
modulus approximation \cite{khach}. With the total free energy
of an inhomogeneous system written as a function of order-parameter
fields, the temporal evolution of microstructures during a phase
transformation can be obtained by solving the coupled Cahn-Hilliard
equation for a conserved field $c$, and the time-dependent
Ginzburg-Landau equation for a non-conserved field $\eta_p$
\cite{cahn,hohenberg}:
\begin{equation}
\frac{\partial c}{\partial t} = M\nabla^2 \left[\frac{\partial f}
{\partial c} - \alpha\nabla^2c \right ]
\end{equation}
\begin{equation}
\frac{\partial \eta_p}{\partial t} = -L_p\frac{\delta {\cal F}_{tot}}
{\delta \eta_p}
\end{equation}
where $M$ is related to atom mobility and $L_p$ is the relaxation
constant associated with the order-parameter $\eta_p$. As the
above equations illustrate, the continuum phase-field methodology
depends on three input energies: (1) bulk free energies of solid
solution and precipitate phases, (2) precipitate-matrix
interfacial free energies, and (3) precipitate/matrix lattice
elastic energies. Experimental determination of these quantities
can be difficult and problematic. Therefore a physically motivated
method for accurately determining these quantities is of critical
importance to predict the microstructure evolution of interest. In
particular, if the quantities can be determined from {\it ab
initio} calculations, the goal of an ``{\it ab initio}'' modeling
of alloy microstructure evolution would be, to a great extent,
achieved \cite{zunger0,defontaine}.

Since direct {\it ab initio} calculations of free energies are
either impractical or impossible with currently available
computational power, a useful method has been developed to extend
the {\it ab initio} energetics to thermodynamic properties of
alloy systems with hundreds of thousands of atoms \cite{wolv},
referred to as the mixed-space cluster expansion (CE). In this
scheme, energetics from {\it ab initio} calculations for a number
of small unit cell ($\sim$ 10 atoms) structures are mapped onto a
generalized Ising-like model for a particular lattice type,
involving 2-, 3-, and 4-body interactions plus coherency strain
energies \cite{zunger}. The Hamiltonian can be incorporated into
mixed-space Monte Carlo simulations of $N \sim 10^5$ atoms,
effectively allowing one to explore the complexity of $2^N$
configurational space. As demonstrated by Vaithyanathan {\it et
al.}, the bulk free energy can be obtained from Monte Carlo
simulations coupled with thermodynamic integration techniques. The
precipitate/matrix interfacial free energies can be determined
from similar Monte Carlo simulations or from low temperature
expansion techniques. The elastic strain energies are of precisely
the same form as the coherency strain energy used to generate the
mixed-space CE. Hence, from a combination of {\it ab initio}
calculations, a mixed-space CE approach, and Monte Carlo
simulations, one can obtain all the driving forces needed as input
to the continuum phase-field model. The incorporation of these
energetic properties, obtained from atomistics, into a continuum
microstructural model represents a bridge between these two length
scales, and opens the path toward predictive modeling of
microstructures and their evolution.

To illustrate the use of the method, we mention briefly the work
of Vaithyanathan {\it et al.} who studied the problem of
precipitation of the $\theta'$ (Al$_2$Cu) phase in Cu-Al alloys.
The free energy of the $\theta'$ phase is obtained from {\it ab
initio} calculations of the energy at $T=0$ K, coupled with the
calculated vibrational entropy of this phase. The bulk free
energies of matrix and precipitate phases are then fit to the
local free energy as a function of order-parameter fields in the
phase-field model. $T=0$ K interfacial energies are determined
from supercell calculations, both for the coherent interface and
for the incoherent interface. The anisotropy of these interfacial
energies is large and has been incorporated in the phase-field
model. Elastic energy calculations for the coherent strain of
Al/Al$_2$Cu ($\theta'$) and the calculated lattice parameters of
each phase determine the elastic driving force in this system.
Having determined all the necessary thermodynamic input,
Vaithyanathan {\it et al.} were able, for the first time, to
clarify the physical contributions responsible for the observed
morphology of $\theta'$ precipitate microstructure.  The agreement
between the calculated and experimentally observed microstructure
of $\theta'$ in the Al-Cu alloys was excellent, confirming the
validity of the approach.

Although the phase-field model is able to predict complex
microstructure evolution during phase transformations, it requires
as input phenomenological thermodynamic and kinetic parameters.
For binary systems, {\it ab initio} calculations can provide these
parameters for the phase-field model, but it is unrealistic to
assume that such calculations can be used to determine all the
thermodynamic information for systems beyond ternary. Therefore
semi-empirical methods, such as CALPHAD (calculated phase-diagram)
will remain a useful tool in such an endeavor
\cite{kaufman,saunders,spencer}.

\subsection{Other sequential approaches}

Kinetic Monte Carlo (KMC) simulations, coupled with atomistically
determined kinetic energy barriers, represent a powerful class of
sequential multiscale approaches. For example, a large body of
research has been carried out for surface growth phenomena with
KMC simulations whose kinetic energy barrier parameters for
relevant elemental processes are supplied by {\it ab initio}
calculations \cite{kandel,scheffler}. In an altogether different
field, Cai {\it et al.} have used KMC method to study dislocation
motion in Si based on the well-established double-kink mechanism
\cite{cai}. In their approach, the dislocation is represented by a
connected set of straight line segments which move as the
cumulative effect of a large number of kink nucleation and
migration processes. The rate of these processes is calculated
from transition state theory with the transition energy barrier
having contributions from both atomistically determined energetics
(double-kink formation and migration energy) and elastic
interactions with other dislocation segments as well as from the
externally applied stress.

An example of a multiscale approach, in which KMC is a key
component, employs the so-called level-set method \cite{ratsch,
gyure} for the largest (macroscopic) scale. This approach is
particularly well suited for the study of epitaxial growth, a
subject of great importance in microelectronics and
optoelectronics applications. In the level-set method, growth is
described by creation and subsequent motion of island boundaries.
The model treats the growing film as a continuum in the lateral
direction, but retains atomistic discreteness in the growth
direction.  In the lateral direction, continuum equations
representing the field variables can be coupled to growth through
island evolution, by solving the appropriate boundary-value
problem for the field and using local values of this field to
determine the velocity of the island boundaries. The central idea
behind the level-set method \cite{osher} is that any boundary
curve $\Gamma$, such as a step or the boundary of an island, can
be represented as the set of values $\varphi = 0$ (the level-set)
of a smooth function $\varphi$. For a given boundary velocity {\bf
$v$}, the equation for $\varphi$ is
\begin{equation}
\frac{\partial \varphi}{\partial t} + {\bf v}\cdot \nabla\varphi=0
\label{set}
\end{equation}
Growth is naturally described by the smooth evolution of $\varphi$
determined by this differential equation. In the case of
multilayer growth, the boundaries $\Gamma_k(t)$ of the islands are
defined as the set of spatial points {\bf $x$} for which
$\varphi({\bf x},t) = k$ for $k = 0, 1, 2, \dots$. The evolution
of the level-set function $\varphi$ can be obtained by numerically
solving Eq. (\ref{set}) using non-oscillatory methods \cite{shu}.
The key parameters entering the model are diffusion constants (the
terrace and island-edge diffusion constants) which can in
principle be supplied from atomistic calculations, through the
following procedure (see Fig. \ref{levset}): first, the atomistic
processes are identified which are responsible for terrace or
island-edge diffusion and their energetics are analyzed using
atomistic (possibly {\em ab initio}) calculations; next, the
energy barriers for the atomistic processes are incorporated in a
KMC model which provides the means for coarse-graining the
atomistic degrees of freedom to a few mesoscopic degrees of
freedom describing the evolution of surface features (the island
step edges); finally, the results of the KMC model are
coarse-grained to provide the input to the level-set equations,
that is, they define the values of the boundary velocity {\bf v}
which depends on the local surface morphology. The coarse-graining
between scales eliminates degrees of freedom that are not
essential, making the passage to the next scale feasible. For
example, in the illustration shown in Fig. \ref{levset}, the
smallest step width in the KMC scale corresponds to a two-atom
wide region at the microscopic scale, a situation that is relevant
to the Si(100) surface and possibly to other semiconductor
surfaces (such as III-V compound surfaces). In these cases,
surface atoms tend to be bound to dimer pairs, which is the
essential unit that determines the step structure, even though the
underlying dynamics may be determined by the motion of individual
atoms. Thus, the KMC simulation need only take into account
structures consisting of dimer units, the dynamics of which
determine the step-edge motion needed for the level-set
simulation. The middle terrace in Fig. \ref{levset}(b) is shown as
a grid of squares, each representing a four-atom cluster and being
the minimal unit relevant to step motion at the KMC scale in this
example, assuming that only steps of width equal to two atoms in
each direction are stable. The level-set method is a manifestly
multiscale approach, combining information from three different
regimes (atomistic, mesoscopic and continuum) into a neatly
integrated scheme. Recently, the level-set method has also been
applied to study dislocation dynamics in alloys \cite{srolovitz}.

Yet another sequential multiscale approach has been successfully
applied to the study of crystal plasticity.  This is the DD method
mentioned earlier, incorporating dislocation motion at the
macroscopic scale, the mechanism ultimately responsible for
crystal plasticity. In order to predict the mechanical properties
of materials using DD simulations, a connection between
micro-to-meso scales must be established because dislocation
interactions at close range (when the cores intersect, for
instance), are totally beyond the reach of continuum models. Along
these lines, Bulatov {\it et al.} were able to study dislocation
reactions and plasticity in fcc metals \cite{vasily2} that compare
well with deformation experiments, by integrating the local rules
derived from atomistic simulation of dislocation core interactions
into the DD simulations. The same idea has been further explored
by Rhee {\it et al.} in a study of the stage I stress-strain
behavior of bcc single crystals \cite{rhee}.

\section{Concurrent multiscale approaches}

Broadly speaking, a concurrent multiscale approach is more general
in scope than its sequential counterpart because the concurrent
approach does not rely on any assumptions (in the form of a
particular coarse-graining model) pertaining to a particular
physical problem. As a consequence, a successful concurrent
approach can be used to study many different problems. For
example, dislocation core properties, grain boundary structure and
crack propagation could all be modeled individually or
collectively by the same concurrent approach, as long as it
incorporates all the relevant features at each level. What is
probably most appealing, however, is that a concurrent approach
does not require {\it a priori} knowledge of the physical
quantities or processes of interest. Thus, concurrent approaches
are particularly useful to explore problems about which little is
known at the atomistic level and its connection to larger scales,
and to discover new phenomena.  We discuss below three instances
of concurrent approaches in some detail, and mention some
additional examples more briefly.

\subsection{Macroscopic Atomistic {\it Ab initio} Dynamics}

Fracture dynamics is one of the most challenging problems in
materials science and solid mechanics. Despite nearly a century of
study, several important issues remain unsolved. In particular,
there is little fundamental understanding of the brittle to
ductile transition as a function of temperature in materials;
there is still no definitive explanation of how fracture stress is
transmitted through plastic zones at crack tips; and there is no
complete understanding of the disagreement between theory and
experiment regarding the limiting speed of crack propagation.
These difficulties stem from the fact that fracture phenomena are
governed by processes occurring over a wide range of length scales
that are all connected, and all contribute to the total fracture
energy \cite{needleman2}. In particular, the physics on different
length scales interacts dynamically, therefore a sequential
coupling scheme would not be adequate for the study of fracture
dynamics.

To address these challenges, Abraham, Broughton, Bernstein and
Kaxiras developed a concurrent multiscale modeling approach that
dynamically couples different length scales
\cite{abraham1,abraham2}. This multiscale methodology aims at
linking the length scales ranging from the atomic scale, treated
with a quantum-mechanical tight-binding approximation method,
through the microscale, treated via the classical molecular
dynamics method, and finally to the mesoscale/macroscale treated
via the finite element method in the context of continuum
elasticity. These authors applied this unified approach, termed
macroscopic-atomistic-{\it ab initio} dynamics (MAAD), to the
study of the dynamical fracture process in Si, a typical brittle
material. In traditional studies of fracture, only the continuum
mechanics level (employing, for instance, the FE method) is
usually invoked to account for the macroscopic behavior. But since
there is no natural small-length cutoff present in the continuum
mechanics approach, any important aspect of the atomic-scale
mechanisms for fracture is completely missed. This can be remedied
by introducing classical MD to the simulations. In particular, the
MAAD approach employed the Stillinger-Weber \cite{sw} interatomic
empirical potential for Si to perform MD calculations at the
atomistic level, for a large region of the material near a crack
tip.  However, the treatment of formation and breaking of covalent
bonds at the atomic scale is not reliable with any empirical
potential, because bonds between atoms are an essentially quantum
mechanical phenomenon arising from the sharing of valence
electrons. On the other hand, small deviations from ideal bonding
arrangements can be captured accurately by empirical potentials,
because they are to first approximation harmonic, a feature that
is easily incorporated in empirical descriptions of the
interaction between atoms. Therefore, it was deemed necessary to
include a quantum mechanical approach into the simulations for a
small region in the immediate neighborhood of the crack tip, where
bond breaking is prevalent during fracture, while further away
from this region the empirical potential description is adequate.
The particular methodology chosen to model the immediate
neighborhood of the crack-tip, a semi-empirical nonorthogonal
tight-binding scheme \cite{tb}, describes well the bulk,
amorphous, and surfaces properties of Si. Fig. \ref{maad} shows
the spatial decomposition of the computational cell into five
different dynamic regions of the simulation: the continuum FE
region at the far-field where the atomic displacements and strain
gradients are small; the atomistic MD region around the crack with
large strain gradients but with no bond breaking; the quantum
mechanical region (labelled TB because of the use of the
tight-binding method) right at the crack tip where atomic bonds
are being broken and formed; the FE/MD hand-shaking region; and
the MD/TB hand-shaking region. The total Hamiltonian, $H_{tot}$
for the entire system was written as:
\begin{eqnarray*}
H_{tot}& =& H_{FE}(\{{\bf u,\dot{u} }\}\in FE) + \\
& & H_{MD}(\{ {\bf r,\dot{r}} \} \in MD) + \\
& & H_{TB}(\{ {\bf r,\dot{r}} \} \in TB) + \\
& & H_{FE/MD}(\{ {\bf u,\dot{u},r,\dot{r}} \}
\in FE/MD) + \\
& & H_{MD/TB}(\{ {\bf r,\dot{r}} \} \in MD/TB)
\end{eqnarray*}
The degrees of freedom of the Hamiltonian are atomic positions
{\bf r} and velocities  $\dot{\bf r}$ for the TB and MD regions,
and displacements {\bf u} and their time rates of change $\dot{\bf
u}$ for the FE regions. Equations of motion for all the relevant
variables in the system are obtained by taking appropriate
derivatives of this Hamiltonian. All variables can then be updated
in lock-step as a function of time using the same integrator. Thus
the entire time history of the system may be obtained numerically
given an appropriate set of initial conditions. Following
trajectories dictated by this Hamiltonian leads to evolution of
the system with conserved total energy, which ensures numerical
stability.

The individual approaches at each level (FE, MD and TB) are well
established and tested methods.  What was much more important in
this study was the seamless hand-shaking of the different methods
at the interface of the respective domains, namely the
hand-shaking algorithms between FE and MD regions and between the
MD and TB regions. We present here the main ideas behind the coupling of
the different regions.\\
{\em FE/MD coupling}: To achieve the FE/MD hand-shaking, the FE
mesh spacing is scaled down to atomic dimensions at the interface
of the two regions.  In Fig. \ref{maad}, the FE nodes are
indicated as small open circles connected by thin lines. Moving
away from the FE/MD region and deep into the continuum, one can
expand the mesh size. In this way, the atomistic simulation is
embedded in a large continuum solid, indicated by a green-colored
region in Fig. \ref{maad}(a). FE cells contributing fully to the
overall Hamiltonian (unit weight) are marked with thin solid
lines, while cells contributing to the hand-shake Hamiltonian
(half weight) are represented by thin dashed lines.  Interactions
between the atoms on the MD side, which are represented by an
interatomic potential, carry full weight when fully inside the MD
region (thick solid lines joining neighboring atoms) and half
weight (thick dashed lines) when they cross the boundary, with one
of the neighbors effectively represented by a node in the FE
region. The FE/MD interface is chosen to be far from the fracture
region. Hence, the atoms of the MD region and the displacements of
the FE lattice can be unambiguously mapped onto one another. The
assignment of weights equal to unity within each region and equal
to one half at the interface is arbitrary and can be generalized
by the introduction of a smooth step function.
\\
{\em  MD/TB coupling}: At this interface, the atoms treated
quantum mechanically are shown in red while those treated
classically are shown in green. The dangling bonds at the edge of
the TB region are terminated with pseudo-hydrogen atoms. The
Hamiltonian matrix elements of these pseudo-hydrogen atoms are
carefully constructed to tie off a single Si bond and to ensure
the absence of any charge transfer when that atom is placed in a
position commensurate with the Si lattice. In other words, the TB
terminating atoms are fictitious monovalent atoms forming covalent
bonds with the strength and length of bulk Si bonds.  These
fictitious atoms were called ``silogens'': they behave
mechanically just like Si, but chemically like H. The TB
Hamiltonian including silicon-silicon and silicon-silogen matrix
elements is then diagonalized to obtain electronic energies and
wavefunctions, from which the total energy can be computed. Thus,
at the perimeter of the MD/TB region, there are silogens sitting
directly on top of the atoms of the MD region, which are shown as
the smaller red circles on top of green circles in Fig.
\ref{maad}. On one side of the TB/MD interface, the bonds to an
atom are derived from the TB Hamiltonian, and are shown as shaded
regions in Fig. \ref{maad}, to indicate the electronic
distribution responsible for the formation of the covalent bonds.
On the other side of the interface, the bonds are derived from the
interatomic potential of the MD simulation. The MD atoms of the
interface have a full complement of neighbors, including neighbors
whose positions are determined by the dynamics of atoms in the TB
region; these are shown as small green circles on top of the red
circles in Fig. \ref{maad}. As before, the TB code updates atomic
positions in lock-step with its FE and MD counterparts.

The MAAD approach was employed to study the brittle fracture of Si
in a geometry containing a small crack (notch) within an otherwise
perfect solid, with the exposed notch face in the (100) plane and
the notch pointed in the $\langle$010$\rangle$ direction. The
system consisted of  258,048 mesh points in each FE region,
1,032,192 atoms in the MD region, and approximately 280 unique
atoms in the TB region (for computational reasons, the entire
region modeled by the TB method was broken into smaller, partially
overlapping regions, each assigned to a different processor in a
parallel implementation). The lengths of the MD region are 10.9
\AA~ for the slab thickness along the front of the crack, 3649
\AA~ in the primary direction of propagation, and 521 \AA~ in the
direction of pull (before pulling). Periodic boundary conditions
were imposed at the slab faces normal to the direction of the
crack propagation (along the front of the crack). The wall-clock
time for a TB force update was 1.5 s, that for the MD update was
1.8 s, and that for the FE update was 0.7 s. The TB region was
relocated after every 10 time steps to ensure that it remains at
the very tip of the propagating crack. The computational slab was
initialized at zero temperature, and a constant strain rate was
imposed on the outermost FE boundaries defining the opposing
horizontal faces of the slab. Furthermore, a linear velocity
gradient was applied within the slab, which results in an
increasing internal strain with time. It was observed that the Si
solid failed in brittle fashion at the notch tip when the material
is stretched by $\sim$ 1.5\%. The limiting speed of crack
propagation was found to be 85\% of the Rayleigh speed with the
chosen computational setup. In the course of the simulation, the
straight-ahead brittle cleavage of the Si slab left behind a rough
surface, with increasing roughening as a function of crack
distance. Based on these results, the authors suggested that the
roughening surface is due to the spawning of dislocations with low
mobility on the time-scale of the crack motion.

A general problem associated with domain decomposition, as in the
MAAD simulations, is the spurious reflection of elastic waves
(phonons) at the domain boundaries due to the changes in system
description across the boundaries. For example, such effects have
been observed in the atomistic modeling of dislocation motion
\cite{ohsawa}, crack propagation \cite{gao, zhou,holian,gumbsch},
and energetic particle-solid collisions \cite{carroll,moseler},
all of which involved some domain coupling scheme. Since the MAAD
method involves domain decomposition into the TB, MD and FE
regions, the quality of coupling between different regions needs
to be examined. In a subsequent paper, the same authors reported
that there was no visible reflection of phonons at the FE/MD
interface, and no obvious discontinuities at the MD/TB interface
\cite{abraham2}.  Thus, in this scheme the coupling between the
various domains is indeed performed in a seamless manner, closely
mimicking the actual behavior of the physical system under
investigation. Overall, the MAAD approach represents the state of
the art of current multiscale simulation strategies. It is a
finite-temperature, dynamic and parallel algorithm which, at least
as far as general computational aspects are concerned, is
applicable to any type of material.

Ongoing efforts are exploring the possibility of applying the MAAD
strategy to study chemical effects on mechanical properties of
metallic alloys, such as impurity effects on dislocation motion,
crack nucleation and propagation in various metals. There is an
important qualitative difference between such systems and the
study of brittle fracture of Si mentioned above: the nature of
bonds in metallic systems is very different from the simple
covalent bonds in Si.  This makes necessary the development of a
different way of coupling the quantum mechanical to the classical
atomistic region, because it is no longer feasible to terminate
the bonds at the boundary of the quantum region by simply
saturating them with fictitious atoms like the silogens. In such
endeavors, other more efficient and versatile quantum mechanical
formulations are desirable.  One candidate is the linear scaling
real-space kinetic energy functional method \cite{choly}. This
method approximates the non-interacting kinetic energy of DFT as a
functional of electron density, and electronic
wave-functions are thus eliminated from calculations, and
therefore the method is termed
as orbital-free density functional theory (OFDFT). As a consequence, 
no diagonalization of the electronic Hamiltonian and no sampling of
reciprocal space are necessary, making the method computationally
efficient \cite{carter}.  In particular, the explicit real-space
feature of this approach makes it naturally suitable for domain
coupling within the MAAD framework. While efforts to construct a
fully functioning scheme along these lines are continuing, we
believe this is a promising method with great potential for
applications in metallic systems, which are difficult to handle
with other techniques.

\subsection{Quasicontinuum model}
One observation from many large-scale atomistic simulations is
that only a small subset of atomic degrees of freedom do anything
interesting. The great majority of the atoms behave in a way that
could be described by effective continuum models like elasticity
theory.  The computation and storage of the uninteresting degrees
of freedom - necessary for a fully atomistic calculation - consume
a large proportion of computational resources. This observation
calls for novel multiscale approaches which can reduce the number
of degrees of freedom in atomic simulations
\cite{kohlhoff,gumbsch2}. One such approach proposed by Tadmor,
Ortiz and Phillips is particularly promising and has yielded
considerable success in many applications \cite{tadmor1}. This
concurrent multiscale approach is called the quasicontinuum
method, which seamlessly couples the atomistic and continuum
realms. The chief objective of the approach is to systematically
coarsen the atomistic description by the judicious introduction of
kinematic constraints. These kinematic constraints are selected
and designed so as to preserve full atomistic resolution where
required - for example, in the vicinity of lattice defects - and
to treat collectively large numbers of atoms in regions where the
deformation field varies slowly on the scale of the lattice.
Variants of the quasicontinuum model have been developed and
applied in different situations \cite{tadmor1,tadmor2,
shenoy1,miller1,miller2,rodney,shenoy2,tadmor3,shenoy3,smith1,
knap,smith2}. The essential building blocks of the static
quasicontinuum model are: (1) the constrained minimization of the
atomistic energy of the solid; (2) the use of summation rules to
compute the effective equilibrium equations; and (3) the use of
adaptation criteria in order to tailor the computational mesh to
the structure of the deformation field. An extension of the method
to finite-temperature has also been proposed
\cite{shenoy4}.

The quasicontinuum model\footnote{A web site with useful
information related to the quasicontinuum method can be found at
http://www.qcmethod.com, where the quasicontinuum codes are also
available to download.} starts from a conventional atomistic
description, which computes the energy of the solid as a function
of the atomic positions. The configuration space of the solid is
then reduced to a subset of representative atoms. The positions of
the remaining atoms are obtained by piecewise linear
interpolations of the representative atom coordinates, much in the
same manner as displacement fields are constructed in the FE
method. The effective equilibrium equations are then obtained by
minimizing the potential energy of the solid over the reduced
configuration space. A direct calculation of the total energy in
principle requires the evaluation of sums that are extended over
the full collection of atoms, namely,
\begin{equation}
\label{qc1}
E_{tot} = \sum_{i=1}^{N}E_i,
\end{equation}
where $N$ is the total number of atoms in the solid. The full sums
may be avoided by the introduction of approximate summation rules.
For example, the lattice quadrature analog of Eq. (\ref{qc1}) can
be written as
\begin{equation}
\label{qc2}
E_{tot} \approx \sum_{i=1}^{N_r}n_i \bar{E}_i,
\end{equation}
where $n_i$ is the quadrature weight that signifies how many atoms
a given representative atom stands for in the description of the
total energy, and $\bar{E}_i$ is the energy of $i$-th
representative atom. Note that in this case the sum is over the
$N_r$ representative atoms only. In the quasicontinuum approach,
the FE method serves as the numerical tool for determining the
displacement fields, while an atomistic calculation is used to
determine the energy of a given displacement field. The positions
of the coarse-grained atoms are needed because the energy of the
representative atoms depends on them.  This approach is in
contrast to standard FE schemes, where the constitutive law is
introduced through a phenomenological model. The selection of the
representative atoms may be based on the local variation of the
deformation field. For example, near dislocation cores and on
planes undergoing slip, full atomistic resolution is attained with
adapted meshing. Far from defects or other highly stressed
regions, the density of representative atoms rapidly decreases,
and the collective motion of very large numbers of atoms is
dictated, without appreciable loss of accuracy, by a small number
of representative atoms.

The quasicontinuum method has been applied to a variety of
problems, including dislocation structures \cite{tadmor1,tadmor2},
interactions of cracks with grain boundaries \cite{shenoy1},
nanoindentations \cite{tadmor3,smith1,smith2}, dislocation
junctions \cite{rodney}, atomistic scale fracture process
\cite{miller1}, etc. By way of example, Shenoy {\it et al.}
applied the method to study the interaction of dislocations with
grain boundaries (GB) in Al \cite{shenoy1}. In particular, they
considered a reformulation of the quasicontinuum model that allows
for the treatment of interfaces, and therefore of polycrystalline
solids. As the first test of the model, they computed the GB
energy and atomic structure for various symmetric tilt GB's in Au,
Al, and Cu using both direct atomistic calculations and the model
calculations. They found  excellent agreement between the two sets
of calculations, indicating the reliability of the model for their
purpose. In the study of Al, they used nanoindentation-induced
dislocations to probe the interaction between dislocations and
GB's. Specifically, they considered a block oriented so that the
(111) planes are positioned to allow for the emergence of
dislocations which then travel to the $\Sigma$ 21($\bar{2}$41) GB
located at $\sim$ 200 \AA~ beneath the surface [see Fig.
\ref{GB}(a)]. First, the energy minimization is performed to
obtain the equilibrium configuration of the GB, then a mesh is
constructed accordingly as shown in Fig. \ref{GB}(a). The region
that is expected to participate in the dislocation-GB interaction
is meshed with full atomistic resolution, while in the far fields
the mesh is coarser. The displacement boundary conditions at the
indentation surface are then imposed onto this model structure,
and after the critical displacement level is reached, dislocations
are nucleated at the surface.  With the EAM potential \cite{eam}
supplying the atomistic energies in the quasicontinuum approach,
they observed closely spaced (15 \AA) Shockley partials nucleated
at the free surface. As seen from Fig. \ref{GB}(b), the partials
are subsequently absorbed at the GB with the creation of a step at
the GB and no evidence of slip transmission into the adjacent
grain is observed. The resultant structure can be understood based
on the concept of displacement shift complete lattice \cite{dsc}
associated with this symmetric tilt GB. As the load is increased,
the second pair of Shockley partials is nucleated.  These partials
are not immediately absorbed into the GB, but instead form a
pile-up [Fig. \ref{GB}(b)]. The dislocations are not absorbed
until a much higher load level is reached. Even after the second
set of partial dislocations is absorbed at the GB, there is no
evidence of slip transmission into the adjacent grain, although
the GB becomes much less ordered. The authors argued that their
results give hints for the general mechanism governing the
absorption and transmission of dislocations at GB's.  The same
work also studied the interaction between a brittle crack and a GB
and observed stress-induced GB motion (the combination of GB
sliding and migration). In this example, the reduction in the
computational effort associated with the quasicontinuum thinning
of degrees of freedom is significant. For example, the number of
degrees of freedom associated with the mesh of Fig. \ref{GB}(a) is
about 10$^4$, three orders of magnitude smaller than what would be
required by a full atomistic simulation (10$^7$ degrees of
freedom).

Recently, the quasicontinuum model has been extended to complex
Bravais lattices \cite{tadmor4} whereby more complicated materials
can be handled \cite{smith2}. But because of the expression for
the total energy adopted in Eq. (\ref{qc1}), (\ref{qc2}), the
actual atomistic methods that can be implemented in the
quasicontinuum model are limited to ones that can be easily cast
in such a form, if one insists on having the ability to resolve
the FE nodes all the way to the atomic scale.  This limit is often
referred to in the literature as the ``non-local'' regime of the
quasicontinuum method.  In contrast, the ``local'' limit refers to
the case where each FE node represents a very large number of
atomistic degrees of freedom, which is modeled as corresponding to
an infinite solid homogeneously deformed according to the local
strain at the node.  In this limit, it is advantageous to employ  
effective Hamiltonians to compute energetics for the
representative atoms. Such Hamiltonians can be constructed from
{\it ab initio} calculations, and may include physics that
atomistic simulations based on classical interatomic potentials
(such as EAM) are not able to capture. For example, by
constructing an effective Hamiltonian parametrized from {\it ab
initio} calculations, Tadmor {\it et al.} were able to study
polarization switching in piezoelectric material PbTiO$_3$ in the
context of the quasicontinuum model in the local limit\cite{tadmor5}. 
This particular Hamiltonian includes the following terms:
the elastic energy of the lattice, the
coupling between strain and atomic displacement, harmonic and
anharmonic phonon energy contributions, the interaction of atomic
displacement with the applied electric field, and the electrostatic
energy. With this effective Hamiltonian, it was shown that the
behavior of a large-strain ferroelectric actuator under the
application of competing external stress and electric fields can
be modeled successfully, reproducing, for example, all the
important features of the experimental strain vs. electric field
curve for the actuator. The advantage of these simulations is that
they also provide insight into the microscopic mechanisms
responsible for the macroscopic behavior, making possible the
improvement of design of the technologically useful materials.

One pitfall of the quasicontinuum model is the so-called ``ghost
force'' at the interface between the local region, identified with
slow variation of the deformation gradient, and the nonlocal
region, identified with rapid variation of the deformation
gradient \cite{shenoy2}. The error arises from the discontinuity
between the neighboring cells where the cell sizes are less than
the range of the atomistic potential. Care must be taken to
correct these ``ghost forces'' \cite{shenoy2}. Finally we should
point out that the quasicontinuum approach also shares certain
features with sequential approaches, namely, the constitutive
equation for the FE nodes is drawn from atomistic calculations
(akin to message passing in sequential approaches). The reason we
categorize it as a concurrent multiscale approach is that the
atomistic and FE calculations are performed concurrently rather
than in sequence, because the range of deformations encountered in
various parts of the system are not know beforehand.  Moreover,
some sort of domain partitioning (meshing) is involved in the
quasicontinuum approach.

\subsection{Coarse-grained molecular dynamics}
Mesoscopic elastic systems, and in particular
micro-electro-mechanical-systems (MEMS), recently have captured a
great deal of attention and research interest as micro-machines
and devices. However, there is serious concern regarding their
mechanical integrity and stability in applications because these
sub-micron devices are so minuscule that structural defects and
surface effects could have large impact on their performance. On
the other hand, the computational study of the mechanical
properties of the MEMS has turned out to be extremely difficult
because they are too small in size for finite-element simulations
(at the limit where
continuum elasticity theory may be no longer valid), but too
large for atomistic simulations \cite{rudd,rudd1}. To resolve this
problem, a concurrent multiscale simulation strategy called
coarse-grained molecular dynamics (CGMD) has been developed by
Rudd and Broughton \cite{rudd,rudd1}. This approach bears some
resemblance with the quasicontinuum model, yet there exist
important differences between the two.

The CGMD approach is based on a statistical coarse-graining
prescription.  In particular, the model aims at constructing
scale-dependent constitutive equations for different regions in a
material. In general, the material of interest can be partitioned
into cells, whose size varies so that in important regions a mesh
node is assigned to each equilibrium atomic position, whereas in
other regions the cells contain many atoms and the nodes need not
coincide with atomic sites. The CGMD approach produces equations
of motion for a mean displacement field defined at the nodes by
defining a conserved energy functional for the coarse-grained
system as a constrained ensemble average of the atomistic energy
under fixed thermodynamic conditions. The key point of this
effective model is that the equations of motion for the nodal
(mean) fields are {\it not} derived from the continuum model, but
from the underlying atomistic model. The nodal fields represent
the average properties of the corresponding atoms, and equations
of motion (in this particular case Hamilton's equations) are
constructed to describe the mean behavior of underlying atoms that
have been integrated out.

One important underlying principle of CGMD is that the classical
ensemble must obey the constraint that the position and momenta of
atoms are consistent with the mean displacement and momentum
fields. To be specific, let the displacement of atom $\mu$ be
${\bf u}_\mu$ = ${\bf x}_\mu$ - ${\bf x}_{\mu0}$ where ${\bf
x}_{\mu0}$ is its equilibrium position. The displacement of mesh
node $j$ is an average of the atomic displacements
\begin{equation}
{\bf u_j} = \sum_{\mu}f_{j\mu}{\bf u}_{\mu},
\label{constraint}
\end{equation}
where $f_{j\mu}$ is a weighting function, a microscopic analog of
the FE interpolating functions. Note that Latin indices, $j, k$,
denote mesh nodes and Greek indices, $\mu, \nu$, denote atoms. A
similar relation holds for the momenta {\bf p$_{\mu}$}. Since the
nodal displacements are fewer or equal to the atomic positions in
number, fixing the nodal displacements and momenta does not
necessarily determine the atomic positions entirely. Therefore
some subspace of phase space remains not sampled, which
corresponds to the degrees of freedom that are missing from the
mesh points. The coarse-grained energy is defined as the average
energy of the canonical ensemble on this constrained phase space:
\begin{eqnarray*}
E({\bf u}_k,\dot{\bf u}_k)& =& \langle H_{MD}\rangle_{{\bf u}_k,
\dot{\bf u}_k} \\
& & =\int d{\bf x}_\mu d{\bf p}_\mu H_{MD}e^{-\beta H_{MD}}\Delta/Z,
\end{eqnarray*}
\begin{equation}
\Delta = \prod\limits_{j}\delta\left({\bf u}_j -\sum\limits_{\mu}{\bf u}_{\mu}
f_{j\mu}\right)
\delta \left(\dot{{\bf u}}_j -\sum\limits_{\mu}\frac{{\bf p}_{\mu}
f_{j\mu}}{m_\mu} \right),
\end{equation}
where $\beta=1/(k_BT)$ is the inverse temperature and $Z$ is the
partition function.  The 3-D delta functions $\delta({\bf u})$
enforce the mean field constraint [Eq. (\ref{constraint})].

When the mesh nodes and the atomic sites are identical, the CGMD
equations of motion agree with the atomistic equations of motion.
As the mesh size increases some short-wavelength degrees of
freedom are not supported by the coarse mesh. But these degrees of
freedom are not neglected entirely, because their thermodynamic
average effect has been retained. This approximation is expected
to be good if the system is initially in thermal equilibrium, and
the missing degrees of freedom only produce adiabatic changes to
the system. The Hamiltonian was derived originally for a monoatomic
harmonic solid, but can be easily generalized to polyatomic solids
\cite{rudd}. After deriving the equations of motion from the
assumed Hamiltonian for a particular system, one can perform the
CGMD for the nodal points.

As a proof of principle, the method was applied to one-dimensional
chains of atoms with periodic boundary conditions where it was
shown that the CGMD gives better results for the phonon spectrum
of the model system compared to two different FE methods
\cite{rudd}. A variety of other calculations have also been
performed with the CGMD to validate its effectiveness
\cite{rudd1,rudd2}.

Although the CGMD has proven to be reliable in the description of
lattice statics and dynamics, the implementation of the model
varies from system to system. This is because different
approximations have to be made to the Hamiltonian that represents
a particular system. On the other hand, such approximations can be
estimated and controlled in the CGMD method. This advantage makes
the CGMD method a good candidate for replacing the FE method in
the MAAD approach when a high quality of FE/MD coupling is
required. As we alluded earlier, the CGMD approach resembles the
quasicontinuum model in the sense that both approaches adopt an
effective field model, an idea rooted in the renormalization group
theory for critical phenomena. In both approaches, less important
(long wave-length) degrees of freedom are removed while the
effective Hamiltonian is derived from the underlying fine-scale
(atomistic) model. Although both approaches are developed to
couple FE and atomistic models, the quasicontinuum method is
mainly applicable to zero temperature calculations while the CGMD
is designed for finite temperature dynamics. The quasicontinuum
model has shown its success in many applications, but the CGMD
approach has yet to show its wider applicability and versatility.

\subsection{Other works}
Recently a more general model for the dynamics of coarse-grained
multiscale systems was proposed by Curtarolo and Ceder
\cite{curtarolo}. The model is similar to the Migdal-Kadanoff
approach in the renormalization group theory \cite{migdal}, where
the system is coarse-grained through a bond moving process. The
new potentials are constructed to assure that the partial
partition function of the system remains unchanged. The
information removed from the coarse-graining process can be
quantified by the entropy contribution of each step. Although the
model is shown to produce excellent results for mechanical and
thermodynamical properties compared to the non-coarse-grained
system, so far it is limited to two dimensional systems, and its
generalization to three dimensions is yet to be achieved and
tested.

Another interesting approach has been developed by Shilkrot,
Miller and Curtin aimed at linking an atomistic region to a
``defected'' dislocation dynamics region \cite{bill}. In this
coupled atomistic and discrete dislocation (CADD) method, the
fully atomistic region is directly coupled to a linear elastic
continuum region containing dislocations which are modeled as
continuum elastic line defects. The dislocations at the continuum
region are treated with the standard discrete dislocation method
\cite{dd}, and the atomistic region may have any kind of atomic
scale defects. The key aspect of the CADD method is that the
dislocations can pass between the atomistic and continuum regions
smoothly. Two developments have been made to achieve this goal:\\
(1) detection of the dislocation near the atomic/continuum
interface; and\\
(2) a procedure for moving the ``right'' dislocations across the
interface.\\
So far, this approach has only been implemented in 2D systems, but
it has been shown to agree quite well with the 2D atomistic
calculations for Al.

Some other concurrent approaches are similar to the MAAD method
but concentrate on linking two different length scales rather than
three. For example, Bernstein and Hess \cite{noam} have simulated
brittle fracture of Si by dynamically coupling empirical-potential
MD and semi-classical TB approaches. In a similar vein, Lidorikis
{\it et al.} have studied stress distribution in Si/Si$_3$N$_4$
using a hybrid MD and FE approach \cite{lid}. More recently, a
first-principles Green's function boundary condition method has
been developed to self-consistently couple the strain field
produced by a line defect to the long range elastic field of the
host lattice \cite{woodward}.

Concurrent multiscale ideas have also been applied to the
modeling of biomolecules. In particular, the hybrid quantum
mechanical and molecular mechanical (QM/MM) methods have been
gaining ground in the study of proteins and enzymes in which the
small part of a molecule (active site) is modeled by {\it ab
initio} methods while the rest of the molecule can be dealt with
by a more approximate classical force field theory \cite{qm}. One
particular implementation \cite{cui} of the QM/MM strategy is to
combine the quantum mechanical self-consistent-charge
density-functional-based TB method \cite{dftb} with the CHARMM
molecular force fields \cite{charmm}. This approach has been used
to study the reactions catalyzed by triosephosphate isomerase and
the dynamics of small peptide helices in water \cite{cui}. 

Finally
we wish to comment that many concurrent models such as 
the ones 
discussed above are designed for covalently-bonded systems.
These methods take advantage of the localized electron bonding 
across the domain interface (between TB/MD and between QM/MM),
and partition
the bonding energy approximately with a certain degree of empiricism. 
But for metallic systems, the 
bonds are not localized or associated with a distinct pair of atoms,
therefore the concept of ``bonding energy partition'' across the
domain interface becomes invalid, and new concepts are needed.
Recently several groups have exploited the idea of ``embedding 
potential'' in simulations where a region (I) with 
more accurate description of the physics is
embedded into another region (II) with less accurate description.
The influence of region II on region I is described by the
``embedding potential'' which corresponds to a local one-electron
operator in the framework of the DFT \cite{govind,govind1,wesolowski,kluner}.
For example, in an effort to improve the LDA/DFT 
description of molecular adsorption on surfaces, a coupling method 
was developed where
a more accurate (quantum chemical) region (I) was embedded in a 
less accurate LDA/DFT 
region (II)
\cite{govind,govind1,kluner}. The ``embedding potential'' is defined as
the functional derivative of the coupling energy with respect to the
electron density $\rho_I({\bf r})$ in region I. The total
electron density                     
$\rho_{tot}$ =
$\rho_I$ + $\rho_{II}$, where $\rho_{II}$ is the electron density in the LDA/DFT
region, can be obtained by just LDA/DFT calculations for the
entire system since the electron density is usually well represented
by LDA/DFT. 
$\rho_{tot}$ is then held fixed during the subsequent calculations.
By employing the OFDFT method
\cite{carter} for the coupling energy, the ``embedding potential'' can be
explicitly evaluated for any given $\rho_I({\bf r})$. The ``embedding
potential'' as an effective local one-electron operator can in turn be 
added to the Hamiltonian of region I, and the new electron 
density $\rho_I({\bf r})$
is thus determined. In this way, $\rho_I({\bf r})$ can be updated
self-consistently for the given   
$\rho_{tot}$. The same ``embedding potential'' idea can be applied to
the coupling between two different DFT regions,
or between two regions where one is treated with DFT, and 
the other is treated classically, for instance with EAM.

This last approach, currently under development in our
group, deserves some elaboration. This approach strives to combine 
quantum mechanics via OFDFT, classical mechanics via EAM, and continuum
mechanics via the quasicontinuum method in a unified description for 
{\it metallic}
systems. Since the electron density defined in the EAM potential
along with the EAM nuclei, 
could generate an ``embedded potential'' that the OFDFT electrons 
experience, the coupling energy between OFDFT and EAM regions can 
be explicitly
calculated. Furthermore, the EAM atomistic region can be easily
coupled to the continuum region based on the nonlocal description
of the quasicontinuum framework. 

\section{Extending time-scales}
\subsection{Accelerated dynamics
\label{Accel_dyn}} As we have seen, MD plays a critical role in
modeling of materials problems because MD simulations can follow
the actual dynamical evolution of the system without assuming any
mechanism or pathway for the dynamics, in contrast to, say, MC or
molecular statics simulations. However, MD is typically limited to
a time-scale of nanoseconds because standard MD simulations follow
the individual vibrations of all the atoms, whose vibration
frequencies are of the order of 10$^{14}$ s$^{-1}$. This is
particularly troublesome for the complex systems whose dynamics is
characterized by the occurrence of rare but important events, such
as chemical reactions, diffusion processes, and conformational
changes. In these systems, the existence of energetic barriers
much larger than $k_{B}T$ that separate the initial from final state,
leads to reaction times far greater than those that can be
currently accessed computationally. The other reason for extending
time-scales is that time is a sequential object, and the current
progress in parallel computing has little impact on solving the
problem. Therefore, algorithms that could address the time-scale
problem could revolutionize the field of computational materials
science and engineering.

In the past few years, significant progress has been made to
accelerate MD simulations. A class of accelerated dynamics
methods, including hyperdynamics, parallel replica dynamics, and
temperature accelerated dynamics, have been developed by Voter and
coworkers \cite{voter,voter1,voter2}. 
Although each method accomplishes this
acceleration in a different way, transition state theory (TST)
provides the common theoretical foundation. TST is an
elegant theory with extensive applications in materials science
\cite{tst1,tst2,tst3,tst4,tst5,vineyard}. In TST theory, a
state-to-state transition rate constant ($K^{TST}$) is
approximated as the flux through a dividing surface in phase space
separating the initial and final states. A common and useful
approximation to TST is the harmonic TST in which one assumes
that the potential energy surface (PES) near the minimum
can be expanded with the harmonic approximation. Thus the
TST rate constant (the flux through the saddle plane) becomes
\begin{equation}
\label{htst1}
K^{HTST} = \nu_0 \exp(-E/k_BT),
\end{equation}
where
\begin{equation}
\label{htst2}
\nu_0 = \frac{\prod\limits_{i}^{3N} \nu_{i}}
{\prod\limits_{i}^{3N-1} \nu_{i}'}.
\end{equation}
Here $E$ is the activation energy (energy difference
between the minimum
and the saddle point), $\nu_{i}$ is the $i$th normal mode
frequency at the minimum, and $\nu_{i}'$ is the
non-imaginary normal model frequency at the saddle point
\cite{vineyard}. The analytic integration over the whole
phase space yields the well-known Arrhenius temperature dependence.
It is worthwhile to point out that although the exponent depends
only on the barrier height, there is no assumption
that the trajectory passes exactly through the saddle point.
For systems where there is no re-crossing of the dividing surface 
and the modes are
truly harmonic, the rate equation (\ref{htst1}) is exact. 
The underlying concept in the accelerated dynamics 
methods is that the system trajectory 
is simulated to find an appropriate pathway for escape from an
energy well by a process which takes place much faster in the simulation than
it would with direct MD.  We provide below an elementary description of 
this concept. 

The general formulation of TST rests on two assumptions
in order to treat infrequent events: \\
(a) it is known in advance what the different equilibrium states of the system
will be; and \\
(b) it is possible to construct a reasonable dividing surface
along the boundaries between initial and final states (or
equivalently, all the saddle points can be identified). \\
Unfortunately the knowledge of states through which a system may
evolve in most cases (especially in complex systems) is 
incomplete. The hyperdynamics method \cite{voter1,voter2} is
designed to accelerate MD simulations without any advance
knowledge of either the location of the dividing surface or the
states through which the system may evolve. Based on TST,
Voter has shown that it is possible to modify the
PES of the system in such a way that a simulation on
this modified surface exhibits the correct relative probabilities
of transitions, but with enhanced overall transition rates for the
system escaping from one equilibrium state to the various 
nearby equilibrium states. The
key of this approach is to construct a bias potential to raise the
energy of the system in regions other than at the dividing
surfaces. Dynamical evolution with the biased potential leads to
accelerated transition from one equilibrium state to another 
equilibrium state, while the elapsed time
is related to statistical properties of the system. More
precisely, the total time advance for a hyperdynamics simulation
after $n$ integration steps is
\begin{equation}
t_{hyper} = \sum_{j=1}^{n}\Delta t_{MD} e^{\Delta V({\bf r}(t_j))/k_BT},
\label{hyper}
\end{equation}
where $\Delta t_{MD}$ is the time advance for a regular MD
trajectory, $\Delta V({\bf r})$ is the bias potential, and $T$ is
the temperature. The overall computational speed-up is given by
the average boost factor ($t_{hyper}$/$t_{MD}$), divided by the
extra computational cost of calculating the bias potential and its
derivatives. The evolution of hyperdynamics from state to
state is correct because the bias potential does not change the
relative TST rates for different escape paths from a given state.
The long-time dynamics of the simulations are exact to the extent
that the dynamical corrections to the TST are negligible.
Recently, Voter has shown that the bias potential and its
derivatives can be computed in {\it O}(N) fashion without ever
constructing the Hessian \cite{voter2}.  Thus, the
implementation of the hyperdynamics method requires only first
derivatives of the interatomic potential, as for normal MD
simulations.

To demonstrate the effectiveness of the method, Voter has studied
the diffusion of a Ag adatom on the Ag (100) surface at 400 K
using an EAM potential for the energetics. 
He found that a 3.7 $\times$ 10$^6$ steps
of hyperdynamics run gave an average boost of 1356, for a total
time of 9.89 $\pm$ 0.5 $\mu$sec. Each hyperdynamics step required
$\sim$ 30 times the computational time of a direct MD step,
therefore the net computational boost was a factor of 45. The rate constants
obtained from the calculations are in agreement with the full
harmonic rate. For a more complex system with a 10-atom Ag cluster
on the Ag (111) surface at 300 K, he achieved an average boost of
8310 with a hyperdynamics run for 221.2 $\mu$sec. With this
approach, one should be able to observe novel diffusion
mechanisms that can not be accessed by normal MD simulations.

In order to take advantage of recent advances in parallel
computation, Voter proposed the so-called parallel replica
dynamics method \cite{voter3} to treat infrequent events.
For a system in which successive transitions are
uncorrelated, running a number of independent MD trajectories in
parallel gives the exact dynamical evolution between the states.
For a system with correlated crossing events, the state-to-state
transition sequence is still correct, but care must be taken to
eliminate or reduce the error associated with the simulation time.
The parallel replica method represents the simplest and most
accurate of the accelerated dynamics techniques, with the only
assumption being that of infrequent events obeying first-order
kinetics. To be more specific, the probability distribution for
the waiting time before the next escape is assumed to be
\begin{equation}
p(t) = K \exp(-Kt),
\label{replica}
\end{equation}
where $K$ is the rate constant for finding the next escape path
from the current state. In a system that exhibits no correlated
crossing events, $K$ is exactly the TST rate constant ($K^{TST}$).
In a more general case, in which correlated crossings occur, $K <
K^{TST}$. For an $N$-atom system in a particular equilibrium state (potential
energy basin), the entire system is replicated on each of $M$
available parallel or distributed processors. After a short
de-phasing stage during which momenta are periodically randomized
to eliminate correlations between replicas, each processor carries
out an independent constant-temperature MD trajectory for the
entire $N$-atom system, thus exploring the phase space within the
particular basin $M$ times faster than a single trajectory would.
Whenever a transition is detected on any processor, all processors
are alerted to stop. The simulation clock is advanced by the
accumulated trajectory time summed over all replicas, i.e., the
total time spent exploring phase space within the basin before the
escape pathway is found. The parallel replica method also
correctly accounts for correlated dynamical events where TST
is no longer valid. This is accomplished by allowing the
trajectory that made the transition to continue on its processor
for a further amount of time $\Delta t_{corr}$ during which
re-crossings or follow-on events may occur. The simulation clock
is then advanced by $\Delta t_{corr}$, the final state is
replicated on all processors, and the whole process is restarted.
This overall procedure then gives exact state-to-state dynamical
evolution because the escape times obey the correct probability
distribution \cite{voter}. With this approach, significant
extensions of MD time-scales can be achieved. For example, in MD
simulations of vacancy diffusion on the Cu(100) surface at 500 K, a
15-processor parallel computer can give a 14-fold increase in
simulation time per wall-clock time. Moreover, the parallel
replica method can be combined with other accelerated dynamics
methods, such as hyperdynamics to give a multiplicative effect in
the MD time-scale gain \cite{voter3}.

In the temperature-accelerated dynamics (TAD) method, the
transition from state to state is accelerated by increasing
temperature \cite{voter4}. The transitions that should not have
occurred at the original temperature are then filtered out. The
TAD method is more approximate than the previous two methods due
to the fact that it relies on the harmonic TST approximation, but
it often gives substantially bigger boost than the hyperdynamics
or the parallel replica dynamics in systems where the
approximation is justified. Consistent with the accelerated
dynamics concept, the trajectory in TAD is allowed to wander on
its own to find each escape path, so that no prior information is
required about the nature of the reaction mechanisms \cite{voter}.
Like hyperdynamics, TAD can also be combined with the parallel
replica method to achieve an even higher acceleration on parallel
computers.

\subsection{Finding transition pathways
\label{Find_saddl}} As stated earlier, the problem of finding
transition pathways for infrequent events between two known
equilibrium (stable or metastable) states is of considerable interest. 
The accelerated
dynamics methods are designed to find the real dynamic pathways
via effective MD simulations. Other methods requiring no preconceived
mechanism or transition state have also been developed to locate
transition pathways. For example, Elber and Karplus \cite{karplus}
developed a method to find the transition pathway by minimizing
the average value of the potential energy along the path rather
than trying to find the path with the lowest barrier. A more
popular approach, similar in spirit to the Elber-Karplus method, is
the so-called nudged elastic band (NEB) approach \cite{neb1,neb_review}, 
which focuses on the global character of the path rather than on local
properties of the PES.
 
The NEB method is based on the ``chain-of-states'' idea where a number
of images (or replicas or ``states'') of the system are connected 
together between the end-point configurations to
trace out a transition pathway \cite{neb1}. If the images are connected with 
springs of zero natural length, one can define the object function
for this so-called $plain$ elastic band (PEB) method in the following:
\begin{equation}
\label{obj_fun}
S^{PEB}(\vec{R}_1,...,\vec{R}_{P-1})=\sum^{P}_{i=0}{\cal V}(\vec{R}_i)
+ \sum_{i=1}^{P}\frac{Pk}{2}(\vec{R}_i-\vec{R}_{i-1})^2,
\end{equation}
where vector $\vec{R}$ represents the coordinate of the system,
$\cal V$ is the potential energy of the system, and $k$ is the spring
constant. The spring is introduced to ensure that the images are evenly 
spaced along the path. One would envision to find the transition pathway
by minimizing the object function in Eqn. (\ref{obj_fun}) with
respect to the intermediate images while keeping the end-point
images, $\vec{R}_0$ and $\vec{R}_P$ fixed. The force acting on 
the image $i$ is
\begin{equation}
\label{force_neb_1}
\vec{F}_i = -\nabla{\cal V}(\vec{R}_i)+\vec{F}^s_i
\end{equation}
where
\begin{equation}
\vec{F}^s_i = k_{i+1}(\vec{R}_{i+1}-\vec{R}_i)- k_i(\vec{R}_i-\vec{R}
_{i-1}).
\end{equation}
However, as demonstrated
by J$\acute{o}$nsson et al., the PEB
method fails to provide the transition pathway in most situations
\cite{neb1}. For example, if $k_i$ is too large, the elastic band
becomes too stiff, the transition path would then ``cut the corner'' 
and thus miss the saddle point region. On the other hand, if 
$k_i$ is small, the elastic band comes closer to the saddle point,
but the images slide down from the energy barrier (avoid the saddle
point), therefore reducing the resolution of the path in the most
critical region. Furthermore, by noticing the analogy between the
object function in the continuum limit and the action of a classical 
particle of unit mass moving on the inverted PES, J$\acute{o}$nsson 
et al. argued that the particle would move through the saddle point
region with a finite velocity affected by the force component 
perpendicular to the curved path. In other words, the images would
deviate from the minimum energy path (MEP). The problem with ``corner
cutting'' is due to the component of the spring force that is 
perpendicular to the path, which tends to pull images off the MEP.
The problem with sliding down results from the component of the
the potential force or of the true force, $\nabla{\cal V}(\vec{R}_i)$ that 
is parallel to the path. The distance between images becomes uneven
so that the net spring force can balance the parallel component of the
true potential force. To cure these problems,
the NEB method projects out the perpendicular component of the spring 
force and the parallel component of the potential force relative to the 
path. The force on image $i$ becomes
\begin{equation}
\label{force_neb_2}
\vec{F}_i^0 = -\nabla{\cal V}(\vec{R}_i)|_{\perp} + \vec{F}_i^s\cdot
\hat{\tau}_{\parallel}\hat{\tau}_{\parallel},
\end{equation} 
where $\hat{\tau}_{\parallel}$ is the unit tangent to the transition
path and $\nabla{\cal V}(\vec{R}_i)|_{\perp} =
\nabla{\cal V}(\vec{R}_i)-\nabla{\cal V}(\vec{R}_i)\cdot
\hat{\tau}_{\parallel}\hat{\tau}_{\parallel}$. These force projections
(``nudging'') decouple the dynamics of the path itself from the discrete
representation of the path. The spring force thus does not interfere
with the relaxation of the images perpendicular to the path, and 
the relaxed configuration of the images satisfies 
$\nabla{\cal V}(\vec{R}_i)|_{\perp} = 0$, i.e., they lie on the MEP.

The implementation of the NEB in a MD program is quite
simple. First, the energy and its gradient are evaluated
for each image in the elastic band using some description of the
energetics ({\it ab initio} or empirical force fields). 
Then for each image, the local tangent to the path is estimated, and
the force defined in Eq. (\ref{force_neb_2}) is evaluated for an
initial guess of the path. The subsequent minimization for the magnitude
of the forces with respect to the coordinates of the intermediate 
images can be carried out with the velocity Verlet algorithm
\cite{anderson}. Recently, several improvements of the original NEB
have been proposed \cite{neb_review,neb2,neb3}.
The NEB method has found a wide range of applications in 
materials problems, including cross-slip of screw dislocations 
in metals \cite{neb_cross}, diffusion and atomic exchange 
processes at metal and semiconductor surfaces 
\cite{neb_diffu1,neb_diffu2}, dissociative adsorption of molecules 
on surfaces \cite{neb_molecule}, and contact formation of metal
tips on surface \cite{neb_tip}.
The NEB method has been implemented in many 
empirical potential and {\it ab initio} atomistic approaches 
\cite{neb1,neb_review}. One drawback of the NEB method is the
difficulty of choosing appropriate spring constants. A large
spring constant requires a small time step in the evolution 
of states, i.e., more images along the path. A small spring constant,
on the other hand, may fail to achieve the desired uniformity of the
images along the path, and hence may reduce the accuracy for
the energy barrier. Furthermore, like other methods in
this category, the NEB method becomes less
efficient or even inapplicable to systems with very rough energy 
landscapes. 
 
Realizing the importance of real dynamical pathways, Chandler and
collaborators have recently proposed methods for statistically
sampling dynamical paths (MC sampling of MD trajectories) that do
not require the assumption of TST or the existence of a single,
well defined transition state or transition path 
\cite{chandler,chandler1}. In
particular, no reaction coordinate is needed to study the dynamics
or kinetics of rare transitions \cite{chandler2} with these
methods. In a sense, the transition path sampling methods are
metaphorically akin to throwing ropes over rough mountain passes
in the dark: ``throwing ropes'' corresponds to shooting short real
dynamical trajectories;  ``in the dark'' implies that the
high-dimensional systems are so complex that it is generally
impossible to visualize the topography of relevant energy
surfaces. Although these methods are extremely powerful for
treating complex systems with rough energy landscapes, they are 
usually computationally
demanding. In particular, their efficiency usually hinges on 
the ability to produce new accepted paths from old ones, 
thus they have found limited applications so far. 

Recently, 
an alternative finite temperature string method was proposed 
which represents transition paths by their intrinsic parameterization
in order to efficiently evolve and sample paths in the path space
\cite{string_T}. The string method performs a constrained sampling of
the equilibrium distribution of the system in hyper-planes normal
to the transition pathways of a coarse-grained potential which
need not be determined beforehand. The collection of the hyper-planes
is parametrized by a string which is updated self-consistently 
until it approximates locally the correct coordinate associated 
with the reaction event. The region in these planes where the 
equilibrium distribution is concentrated determines a transition
tube in configuration space in which a transition takes place
with high probability. The string method naturally overcomes the
spring constant problem in the NEB method owing to the intrinsic
parameterization of the string, and the distribution of the replicas
along the chain is automatically uniform. The method, however, rests 
on the assumption 
that the equilibrium distribution must be localized on the iso-surfaces 
of the reaction coordinate and these iso-surfaces can be locally 
approximated by the hyper-planes. If the effective transition tube is
highly curved in configuration space this approach may have to
be modified. 

A method for efficiently generating classical trajectories with
fixed initial and final boundary conditions has recently attracted
attention because of its conceptual and computational simplicity
\cite{passerone}. The approach developed by Passerone and
Parrinello addresses a very general problem: given an initial and
a final configuration, what are the dynamical paths that connect
them? Given a classical dynamical system described by a set of
coordinates ${\bf q}$, and its trajectory ${\bf q}(t)$ with
boundary conditions ${\bf q}(0) = {\bf q}_A$ and ${\bf q}(\tau) =
{\bf q}_B$ is determined by locating the stationary point of the
action ${\cal S}$:
\begin{equation}
{\cal S} = \int_{0}^{\tau} {\cal L}({\bf q}(t),\dot{\bf q}(t))dt,
\end{equation}
where ${\cal L}$ is the Lagrangian ${\cal L = T - V}$, 
and ${\cal T}$ and ${\cal V}$ are the
kinetic and potential energy, respectively. Following the work of
Gillilan and Wilson \cite{gillilan}, the action ${\cal S}$ can be
discretized as
\begin{equation}
{\cal S} = \sum_{j=0}^{N-1}\Delta \left[\frac{1}{2}\left(\frac{ {\bf q_j
-q_{j+1}}}{\Delta}\right)^2 - {\cal V}({\bf q_j})\right],
\end{equation}
where ${\bf q}_0 = {\bf q}_A$, ${\bf q}_N = {\bf q}_B$; $\Delta =
\tau/N$ is the time interval and the mass is taken as unitary
\cite{passerone}. The stationary solution of this action is the
discretized Euler-Lagrangian equation and the corresponding
trajectory is identical to the well-known Verlet trajectory. The
novel part of this method is to supplement the action with a
penalty function which favors the energy conserving trajectories:
\begin{equation}
\Theta({\bf q}_j, E) = {\cal S} + \mu\sum_{j=0}^{N-1}(E_j - E)^2.
\end{equation}
Here $\mu$ determines the strength with which the energy
conservation is enforced, $E_j$ is the instantaneous energy given
by
$$E_j = ({\bf q}_j - {\bf q}_{j+1})^2/(2\Delta^2) +
{\cal V}({\bf q}_j)$$ and $E$ is its target value. This is motivated by the
fact that the physical trajectories have to conserve total energy.
In all the systems studied, it was found that there exists a
rather large interval of $\mu$ values such that $\Theta$ has a
minimum close to the Verlet trajectories. In order to minimize the
$\Theta$ function more efficiently, Passerone {\it et al.} make
the transformation
\begin{equation}
{\bf q}_j = {\bf q}_A + \frac{j\Delta}{\tau}({\bf q}_B - {\bf q}_A)
+ \sum_{i=1}^{N}{\bf a}_i \sin\left(\pi i \frac{j\Delta}{\tau}\right),
\end{equation}
thus automatically satisfying the boundary conditions. The
advantage of using ${\bf a}_i$ over ${\bf q}_j$ is that ${\bf
a}_i$ has a global character. In practice, $\Theta$ is first
optimized with respect to a relatively small number of ${\bf
a}_i$, thus capturing the global features of the trajectory, and
then higher frequency terms are added. Each time a standard
conjugate gradient algorithm is used to minimize $\Theta$ with
only the evaluation of the forces. The computational scaling is
therefore linear in the number of degrees of freedom rather than
quadratic as in other approaches involving the Hessian matrix.

To illustrate the performance of the method, Passerone {\it et
al.} studied a few simple systems. For a one-dimensional double
well potential, they found that their solutions agree very well
with the Verlet trajectory but without the calculation of the
Hessian matrix. The second example was a minimization of a
trajectory in a two-dimensional configurational space, namely in
the Mueller potential \cite{mueller}. Again they found a
satisfactory result, namely that the trajectories pass exactly
through the saddle point and the overall behavior of the
trajectories is physical. The last example was to look at a
process in which the central atom of a seven-atom two-dimensional
Lennard-Jones cluster migrates to the surface. In this case, they
claim that their calculations reproduce the results from the more
elaborate method that involves transition path sampling procedure
\cite{dellago}. Finally, the authors pointed out that their method
can be easily implemented within the Car-Parrinello MD approach
\cite{car}, offering a powerful tool for the study of problems in
chemistry and materials science. The main advantages of this
method are the fact that it requires only the calculation of the
forces, its numerical stability and the quality of the
trajectories. Furthermore, the method lends itself to very
efficient parallelization, and it can include naturally the
multiple time-scale approaches \cite{zaloj}.

\subsection{Escaping free-energy minima}
For many systems, the free-energy surface (FES) may have multiple
local minima separated by large barriers, therefore the time-scale
that typical MD and MC simulations can reach is severely limited.
Examples of such systems include conformational changes in
solution, protein folding, and many chemical reactions. A large
number of methods have been developed to overcome the problem,
some of which were already mentioned in sections \ref{Accel_dyn}
and \ref{Find_saddl} (see, for instance, the accelerated dynamics
approach \cite{voter}, or the dynamical transition path sampling
method \cite{chandler}).

Very recently, a novel and powerful method for exploring the
properties of multi-dimensional FES of complex systems has been
proposed by Laio and Parrinello \cite{laio}. This method combines
the ideas of coarse-grained dynamics on the FES \cite{kev,weinan2}
with those of adaptive bias potential methods \cite{huber,landau}.
The method allows the system to escape from local minima in the
FES and at the same time achieves a quantitative determination of
the FES through the integrated process. This method assumes that
there exist a finite number of relevant collective coordinates
$s_{i}, i=1,n$ where $n$ is a small number, and the free energy
${\cal F}(s)$ depends on these parameters. The exploration of the
FES is guided by the generalized forces $F^t_i=- {\partial {\cal
F}} / {\partial s^t_i}$. In order to estimate the forces more
efficiently, an ensemble of $P$ replicas of the system is
introduced, each obeying the constraint that the collective
coordinates have a pre-assigned value $s_{i}$ = $s_{i}^t$. The
coarse-grained dynamics of the collective coordinates is defined
as follows:
\begin{equation}
\sigma^{t+1}_i=\sigma^{t}_i+\delta \sigma \frac {\phi_i^t}
{\left| \phi^t \right|},
\label{dyn}
\end{equation}
where $\sigma^t_i=s^t_i / \Delta s_i $ and $\phi_i^t = F_i^t
\Delta s_i $ are the scaled collective coordinates and forces,
respectively. $\Delta s_i$ is the estimated size of the FES well
in the direction $s_i$, $\left| \phi^t \right| $ is the modulus of
the $n$-th dimensional vector $\phi_i^t$ and $\delta\sigma$ is a
dimensionless stepping parameter. After the collective coordinates
are updated using Eq. (\ref{dyn}), a new ensemble of replicas of
the system with values $\sigma^{t+1}_i$ is prepared, and new
forces $F^{t+1}_i$ are calculated for the next iteration. At the
same time the driving forces are evaluated from the microscopic
Hamiltonian in short standard microscopic MD runs. In order to
explore the FES more efficiently, the forces in Eq. (\ref{dyn})
are replaced by a history-dependent term:
\begin{equation}
\phi_i \rightarrow \phi_i - \frac \partial {\partial \sigma_i} W
\sum_{t' \le t} \prod _i e^{ -\frac {|\sigma_i-\sigma^{t'}_i|^2}
{2 {(\delta \sigma)}^2} }
\label{gauss}
\end{equation}
where the height and the width of the Gaussian, $W$ and
$\delta\sigma$, are chosen in order to provide a reasonable
balance between accuracy and efficiency in exploring the FES. The
component of the forces coming from the Gaussian will discourage
the system from revisiting the same spot and encourage an
efficient exploration of the FES.  As the system diffuses though
the FES, the Gaussian potentials accumulate and fill the FES well,
allowing the system to migrate from well to well. After a while
the sum of the Gaussian terms will almost exactly compensate the
underlying FES well \cite{laio}.

A typical example of this behavior can be seen in Fig.
\ref{gauss1} in which a dynamics in Eq. (\ref{dyn}) is used to
explore a one-dimensional PES ${\cal V}(s)$ with three wells. The
dynamics starts from a local minimum that is filled by the
Gaussians in $\sim$ 20 steps. Then the dynamics escapes from the
well from the lowest energy saddle point, filling the second well
in $\sim$ 80 steps. The second highest saddle point is reached in
$\sim$ 160 steps, and the full PES is filled in a total of $\sim$
320 steps. Hence, in the case of this example, since the form of
the potential is known, it can be verified that for large $t$ and
small $\delta\sigma$,
\begin{equation}
 -\sum_{t' \le t} W
e^{ -\frac {|\sigma-\sigma{t'}|^2}{2 {(\delta\sigma)}^2} }  
\rightarrow {\cal V}(s)
\label{FES}
\end{equation}
modulo an additive constant. Laio and Parrinello also suggest that
Eq. (\ref{FES}) holds for a FES, and the free energy can be
estimated from Eq. (\ref{FES}) for large $t$ \cite{laio}. The
efficiency of the method in filling a well in the PES or the FES
can be estimated by the number of Gaussians that are needed to
fill the well. This number is proportional to $ \left(  1 / \delta
\sigma  \right)^{n} $, where $n$ is the dimensionality of the
problem. Hence, the efficiency of the method scales exponentially
with the number of dimensions involved. A judicious choice of
$\Delta s_i $, $W$ and $\delta\sigma$ will ensure the right
balance between accuracy and sampling efficiency, and the optimal
height and width of the Gaussians can be determined by the typical
variations of the FES.

The method has been applied to the study of dissociation of a NaCl
molecule in water and isomerization of alanine dipeptide in water
\cite{laio}. Overall the method is very efficient in exploring the
FES of complex systems if the collective coordinates are chosen
judiciously. In particular, the topology of a FES can usually be
determined by a few coarse-grained dynamics steps using ``large''
Gaussians. Subsequently, the qualitative knowledge of the FES can
be improved using ``smaller'' Gaussians, effectively reducing the
dimensionality of the problem by exploiting the topological
information obtained with ``large'' Gaussians. As we alluded
earlier, the current method assumes that the free energy ${\cal
F}$($s_i$) depends on a small number of collective coordinates
$s_i$. However, it is not always obvious or possible to identify
such collective coordinates for complex systems {\it a priori}. In
the example of isomerization of alanine dipeptide in water, Laio
{\it et al.} chose the dihedral angles $\Phi$ and $\Psi$ as the
collective coordinates to explore the FES. These authors
recognized that the dihedral angles alone do not provide the
complete description of the dialanine isomerization reaction, and
the real reaction coordinates should include the solvent degrees
of freedom. But their results seemed to reproduce the essential
features of the FES, therefore the authors concluded that the
neglected degrees of freedom, although relevant for determining
the reaction coordinates, are associated with small free energy
barriers and are sampled efficiently during the microscopic
dynamics of the dihedral angles $\Phi$ and $\Psi$. Despite the
success of this particular example, identifying a small number of
collective coordinates {\it a priori}, remains challenging within
this approach. Moreover, the exploration of the FES will be more
efficient if an adaptive way of determining the parameters of
Gaussians could be developed.

\subsection{Other methods}
For systems with a natural disparity in inherent time-scales,
various multiple-time-step integration algorithms have been
developed to deal with them more efficiently
\cite{swindoll,teleman,tuckerman}. One well-known example of such
strategy is the Born-Oppenheimer approximation where the electron
motion is separated from that of the nuclei because of the large
disparity between their masses. In general, the separation of
time-scales occurs when some subset of forces present in the
system is much stronger compared to the rest of the forces while
the masses of the constituents are about the same. For example, in
the simulations of the polyatomic liquids with flexible bonds, the
bond vibrations usually occur at a much faster rate than bond
translation and rotations. In such systems, the configuration
space can be divided into fast and slow degrees of freedom with
the force also being separated into fast and slow components. This
separation yields a set of coupled equations of motion for the
evolution of the fast and slow degrees of freedom. Instead of
solving this set of equations simultaneously, multiple-time-step
integration uses a small time step $\delta t$ to advance the fast
degrees of freedom $n$ steps holding the slow variables fixed. The
slow degrees of freedom are then updated using a time step
$n\delta t$. $n$ can be chosen typically between 5 and 10 in MD
simulations of molecules that can be described in this way. 
Furthermore, if an analytic solution of
high frequency motion can be found, this solution can be
incorporated into an integration scheme for the whole system such
that a time step characteristic of the slow degrees of freedom can
be used and the system can be simulated effectively with a much
smaller number of cycles \cite{tuckerman}.

Recently, a method based on optimization of the action functional
was proposed to extend the time-scale of MD simulations by several orders
of magnitude \cite{elber}. In this method, instead of
parameterizing the trajectory as a function of time, the trajectory
is parametrized as a function of length. Instead of solving the
Newton equations in MD simulations, an action term (stochastic
difference equation with respect to time) is optimized. For
activated processes the method eliminates the ``incubation time'',
and has proven to be very efficient in the simulations of
biomolecules. It remains to be seen, however, if the method can be
applied to problems in materials science.

\section{Conclusions}
In this article we have attempted to provide a comprehensive, if
not exhaustive, overview of the current status of multiscale
simulations methods and their applications in materials science.  
We divided the methods that address multiscale problems in the 
spatial regime into sequential and concurrent.

The {\em sequential} multiscale modeling techniques are in general
more efficient computationally, but they depend on {\it a priori}
knowledge of physical quantities of interest, such as the
$\gamma$-surface in the P-N model, the free-energies in the
phase-field model, and atomistic local laws for mesoscopic DD
simulations. The relevance of these quantities to the
coarse-grained models needs to be carefully examined before the
application of the methods. Furthermore, these approaches should
only be pursued when phenomenological theories (such as the P-N
model or the phase-field model) are well established, therefore
the methods are restricted in their range of application. In
particular, these phenomenological models are often associated
with the assumption of locality (both in space and time). The
example of a local approximation in the phase-field model is
embodied in Eq. (\ref{array1}), which assumes that part of the
energetics of an inhomogeneous system can be written in terms of
quantities obtained for homogeneous systems \cite{rob}. Similarly,
in the P-N model, the $\gamma$-energy is assumed to be constant
within $\Delta x$ distance [see Eq. (\ref{gamma})] in order to
evaluate the total misfit energy. The static approximation
(locality in time) for dynamical properties is also widely used in
phenomenological models. The coupling between different scales in
a sequential approach is usually implicit. A successful sequential
simulation depends equally on the reliability of the
phenomenological model and the accuracy of the 
relevant parameters entering the model.

The {\em concurrent} multiscale approaches are
much more complicated and computationally demanding, but they do
not require {\it a priori} knowledge of physical quantities
supplied from distinct, lower scale simulations. Furthermore,
concurrent approaches do not depend on any phenomenological
models, therefore they are of more general applicability. Although
concurrent approaches are more desirable and appealing, the
actual problem to be attacked must be carefully posed in order to
make the method practical. The problems that may arise in a
concurrent approach are usually associated with the partition of
domains in the system. For example, one needs to dynamically track
the domain boundaries in the MAAD simulations and to adapt the FE
meshes in the quasicontinuum simulations, both of which require
additional care and computational resources. More importantly, in
contrast to a sequential method, a ``good'' hand-shaking in a
concurrent approach between different domains is both
challenging and critical. Although some
interesting ideas have been proposed to remedy the problems of 
coupling between different domains, such as the reflection of
phonons at the domain interface \cite{cai2,weinan,weinan_huang}, there is no
general consensus on what a proper coupling of domains is. A general
criterion that measures the quality of hand-shaking between
domains would therefore be desirable. There is plenty of room for
innovative research on the issue of domain coupling.  
General mathematical formulations of multiscale problems, including
error estimation, may 
turn out to be very useful for practical simulations \cite{weinan2,weinan3}.
In our view, a successful concurrent approach usually has to 
satisfy three conditions: \\
(1) Solve the coupled problem (Hamiltonian) accurately and efficiently
by using ideas such as coarse-graining, ``bonding energy partition'' or 
``embedding potential''. \\
(2) The separate models employed in different domains of the system 
ought to be compatible, i.e., the
physical description of the system due to the distinct models should be
as close as possible; \\
(3) At each level, the individual model should provide a  
good description of its assigned domain. \\
We wish to emphasize the importance of the second condition 
which is not in general well recognized. The
first condition usually guarantees a ``smooth'' hand-shaking across
two domains (e.g., the electron density distribution or the 
displacement field
varies smoothly across the interface), but non-physical charge transfer 
and/or atomic relaxation at
the interface could occur if the second condition is not satisfied.
Therefore a ``smooth'' hand-shaking does not constitute a ``good''
hand-shaking, and a successful concurrent approach relies on 
hand-shakings that are both mathematically accurate and physically 
consistent.

In this overview, we have also described a number of approaches
that strive to extend the temporal scale in the modeling and
simulation of material properties. We categorized these approaches
to methods for accelerating the dynamics, methods for finding
transition paths between equilibrium structures, and methods for
escaping free-energy minima. Although these approaches represent
very significant developments in the field, the problem of linking
the time scale of atomic motion and vibrations (of order femtoseconds) to
scales where interesting physical phenomena are typically studied
(microseconds and beyond) is still wide open in many respects.

Because of the tremendous and continuing progress in multiscale
strategies, this review is by no means exhaustive. We hope that we
have conveyed the message that multiscale modeling is a truly
vibrant enterprise of multi-disciplinary nature. It combines the
skills of physicists, materials scientists, chemists, mechanical
and chemical engineers, applied mathematicians and computer
scientists. The marriage of disciplines and the concomitant
dissolution of traditional barriers between them, represent the
true power and embody the great promise of multiscale approaches
for enhancing our understanding of, and our ability to control
complex physical phenomena.


We acknowledge support from Grant No. F49620-99-1-0272 through
U.S. Air Force Office for Scientific Research (Brown University
MURI) and from the Harvard Materials Research Science and
Engineering Center which is funded by the National Science
Foundation. We thank V.B. Shenoy and A. Laio for providing Fig.
\ref{GB} and Fig. \ref{gauss1}. We are grateful to Nick Choly, 
Weinan E, Paul Maragakis and Sidney Yip for useful comments and 
a critical reading 
of the
manuscript.


\begin{figure*}
\includegraphics[width=400pt]{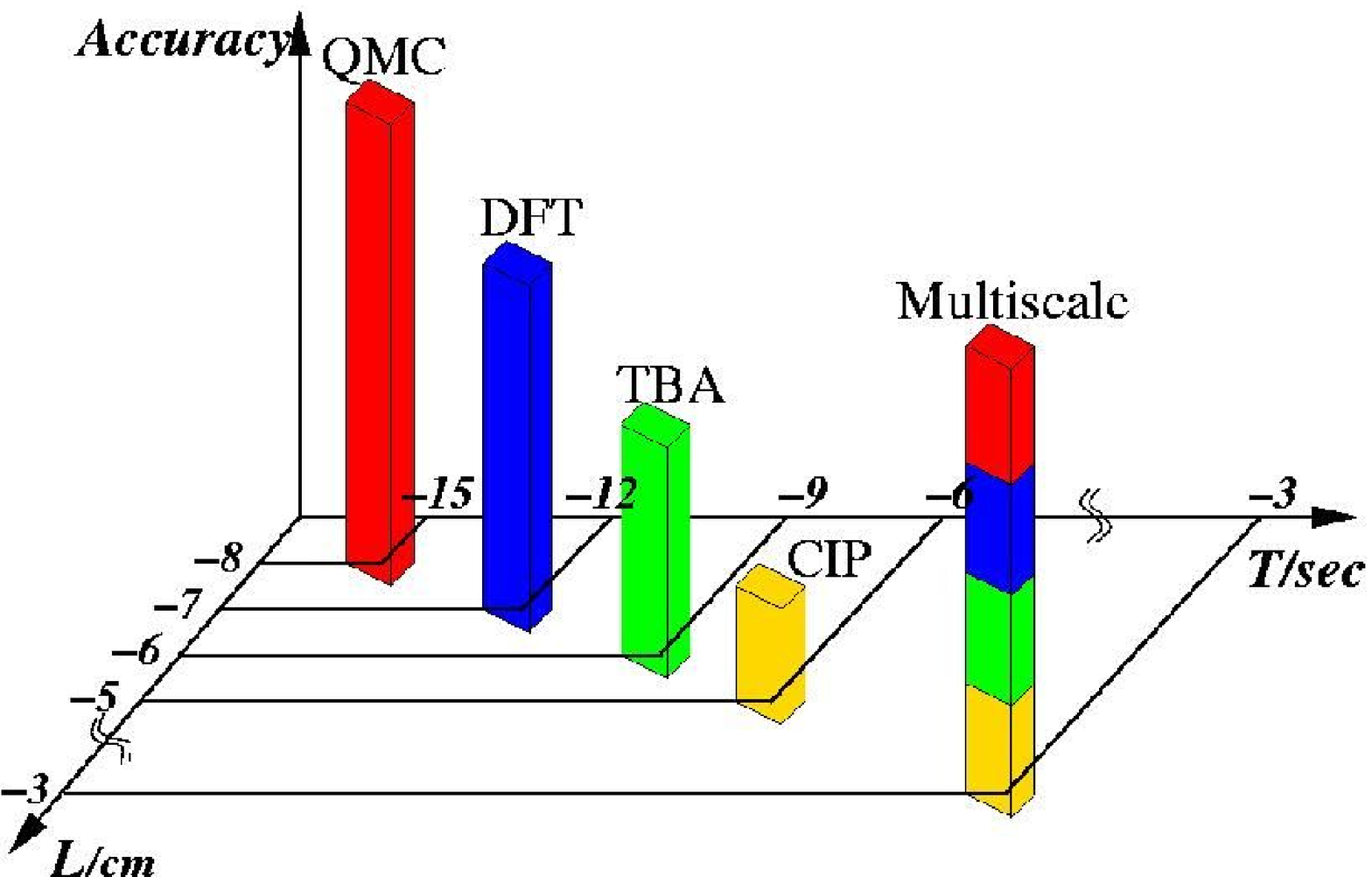}
\caption{A schematic illustration of spacial and temporal scales
achievable by various simulation approaches. The scales are in
centimeters for the length dimension and seconds for the time
dimension, both logarithmic.} \label{scale}
\end{figure*}

\begin{figure*}
\includegraphics[width=400pt]{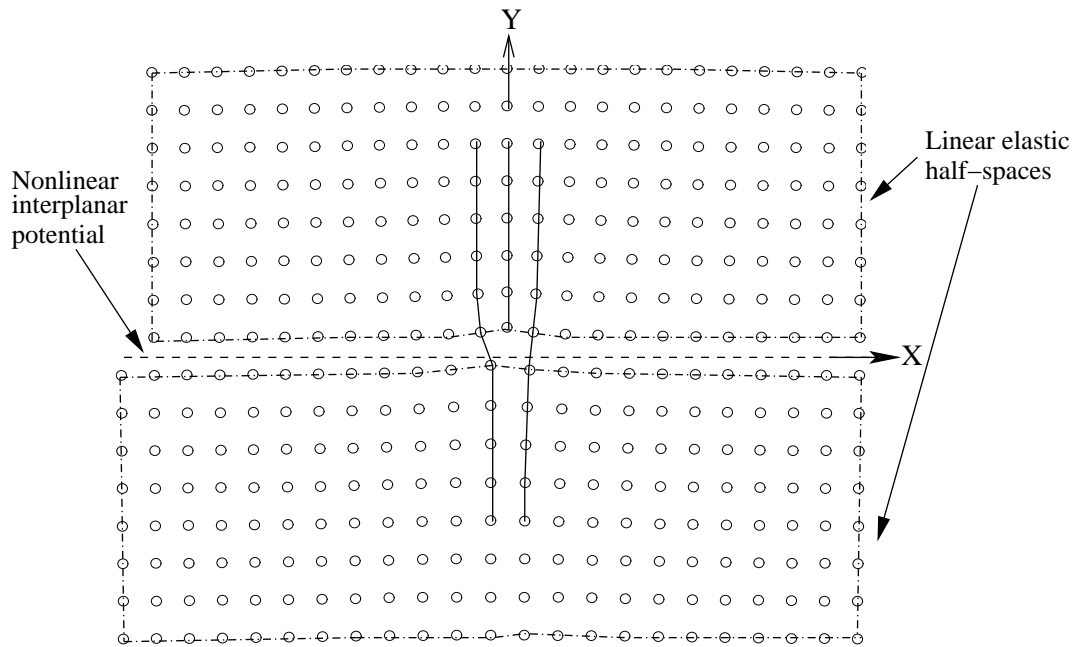}
\caption{A schematic illustration showing an edge dislocation in
a lattice. The partition of the dislocated lattice into linear
elastic region and nonlinear atomistic region allows a multiscale
treatment of the problem.}
\label{edge}
\end{figure*}

\begin{figure*}
\includegraphics[width=400pt]{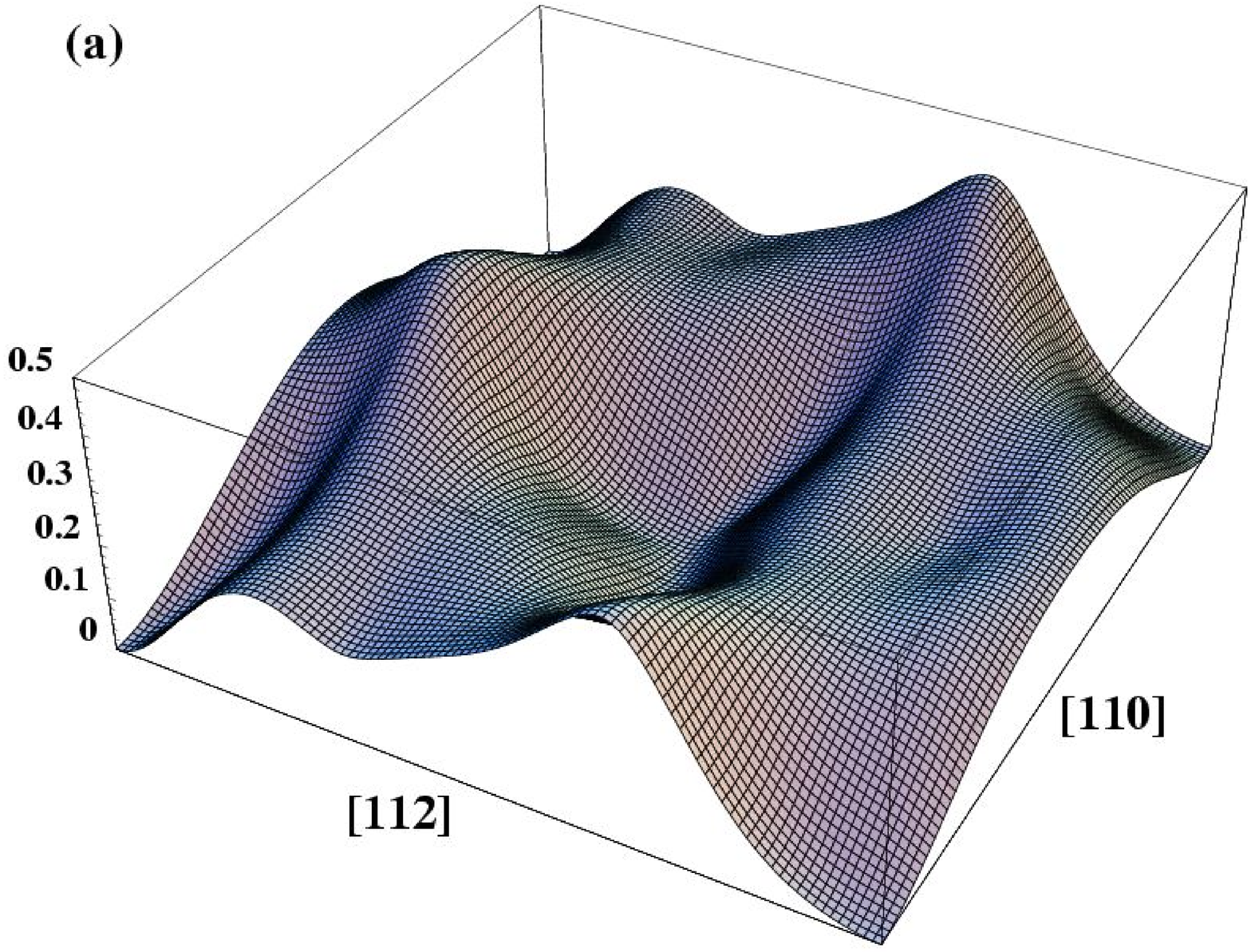}
\includegraphics[width=400pt]{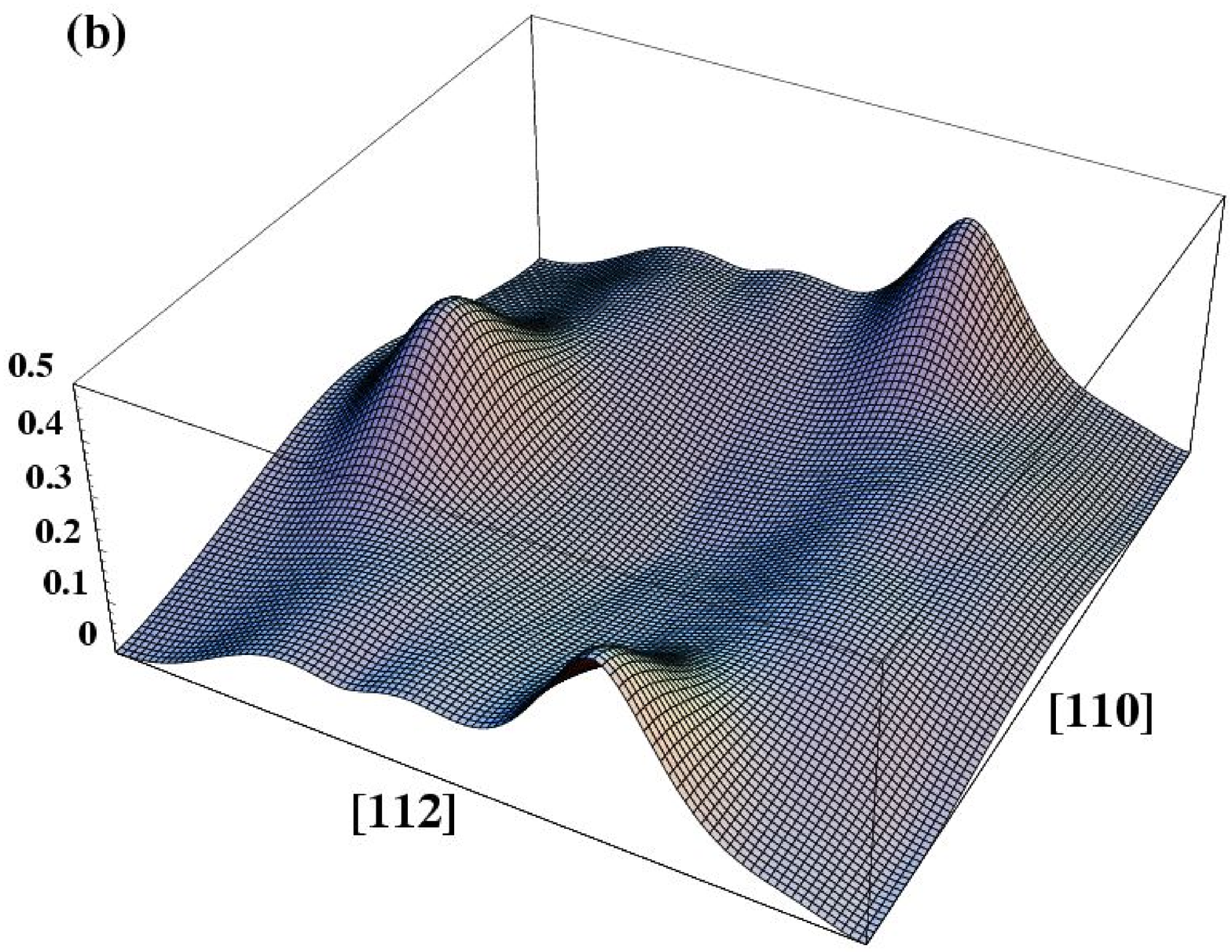}
\caption{The $\gamma$-surface (J/m$^2$) for displacements along a
(111) plane for (a) pure Al and (b) Al+H systems. The corners of
the plane and its center correspond to identical equilibrium
configurations, i.e., the ideal lattice. The two surfaces are
displayed in exactly the same perspective and on the same energy
scale to facilitate comparison of important features.}
\label{gamma2}
\end{figure*}

\begin{figure*}
\includegraphics[width=400pt]{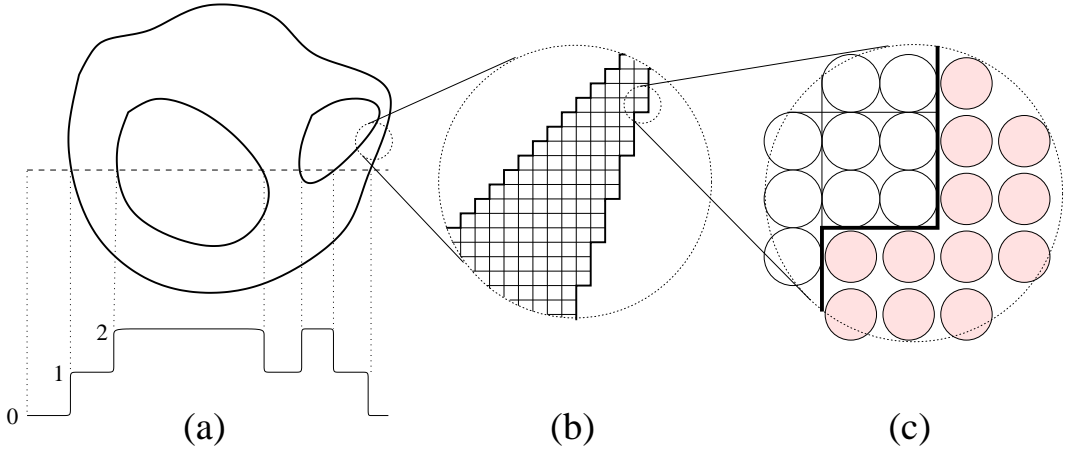}
\caption{Illustration of the three different levels of simulation
in the level-set multiscale approach of surface growth.  (a) The
macroscopic scale, in which island borders are continuous lines
separating heights at different levels (the levels along a
particular cross section are shown schematically, labeled 0, 1 and
2); this scale is treated with the level-set method. (b) The
mesoscopic scale, where the features of the island edges contain
some information about the underlying atomic lattice, indicated
here as the small straight lines that define step directions
consistent with atomic positions; this scale is treated with the
kinetic Monte Carlo approach. (c) The microscopic scale, where the
individual degrees of freedom are explicitly included.  The step
is determined by the positions of atoms in two terraces, the ones
on the upper terrace shown as white larger circles while the ones
on the lower terrace shown as shaded smaller circles; this scale
is treated by atomistic (possibly {\it ab initio}) methods. All
views in this schematic representation are top views (see text for
details).} \label{levset}
\end{figure*}

\begin{figure*}
\includegraphics[width=500pt]{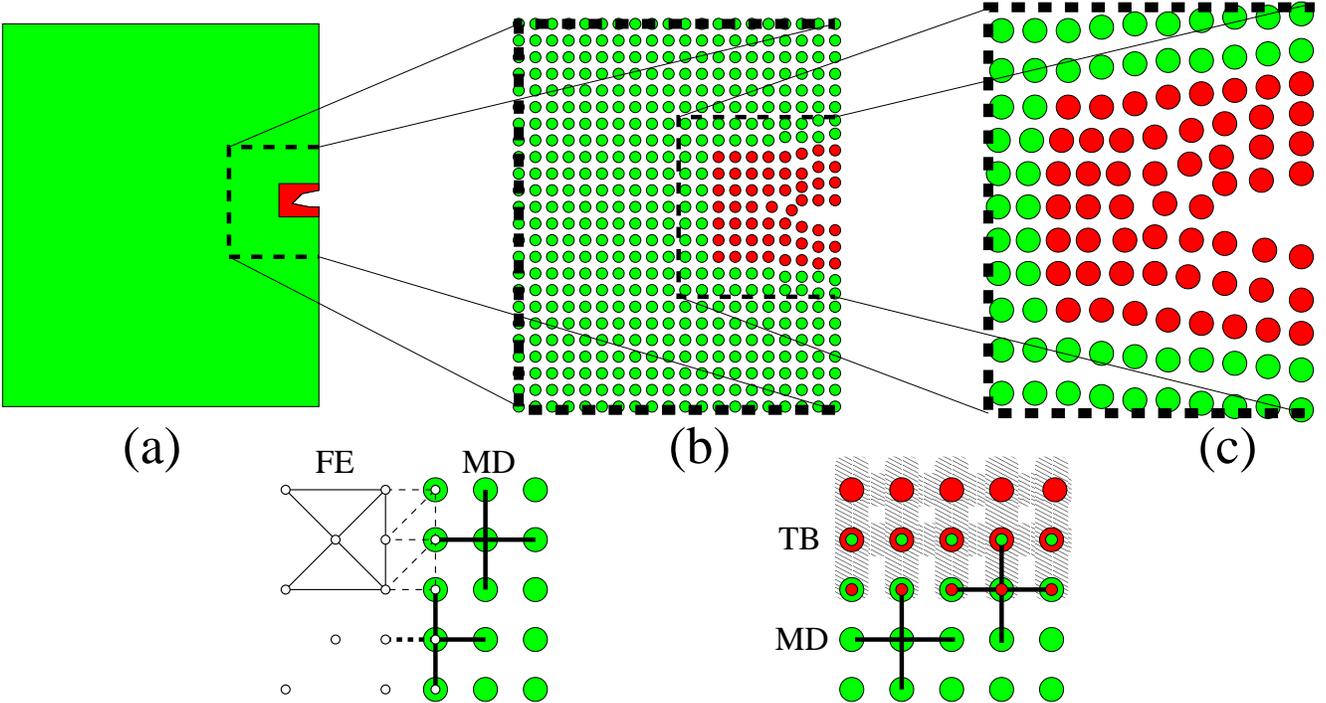}
\caption{Geometrical decomposition of a Si slab with a small crack
into different dynamic regions in the a MAAD simulation: (a) The
system at the macroscopic scale, which is modeled as a continuum
using finite elements (FE), except for the region near the crack
outlined in dashed line. (b) The mesoscopic scale, treated
atomistically with interatomic potentials and molecular dynamics
(MD) with the atoms indicated by green circles, except for the
region in the immediate neighborhood of the crack, outlined in
dashed line. (c) The microcopic scale, treated atomistically with
forces derived from quantum mechanical calculations with the atoms
indicated by red circles. The hand shaking regions between FE and
MD and between MD and TB are also shown schematically (see text
for details). } \label{maad}
\end{figure*}

\begin{figure*}
\includegraphics[width=400pt]{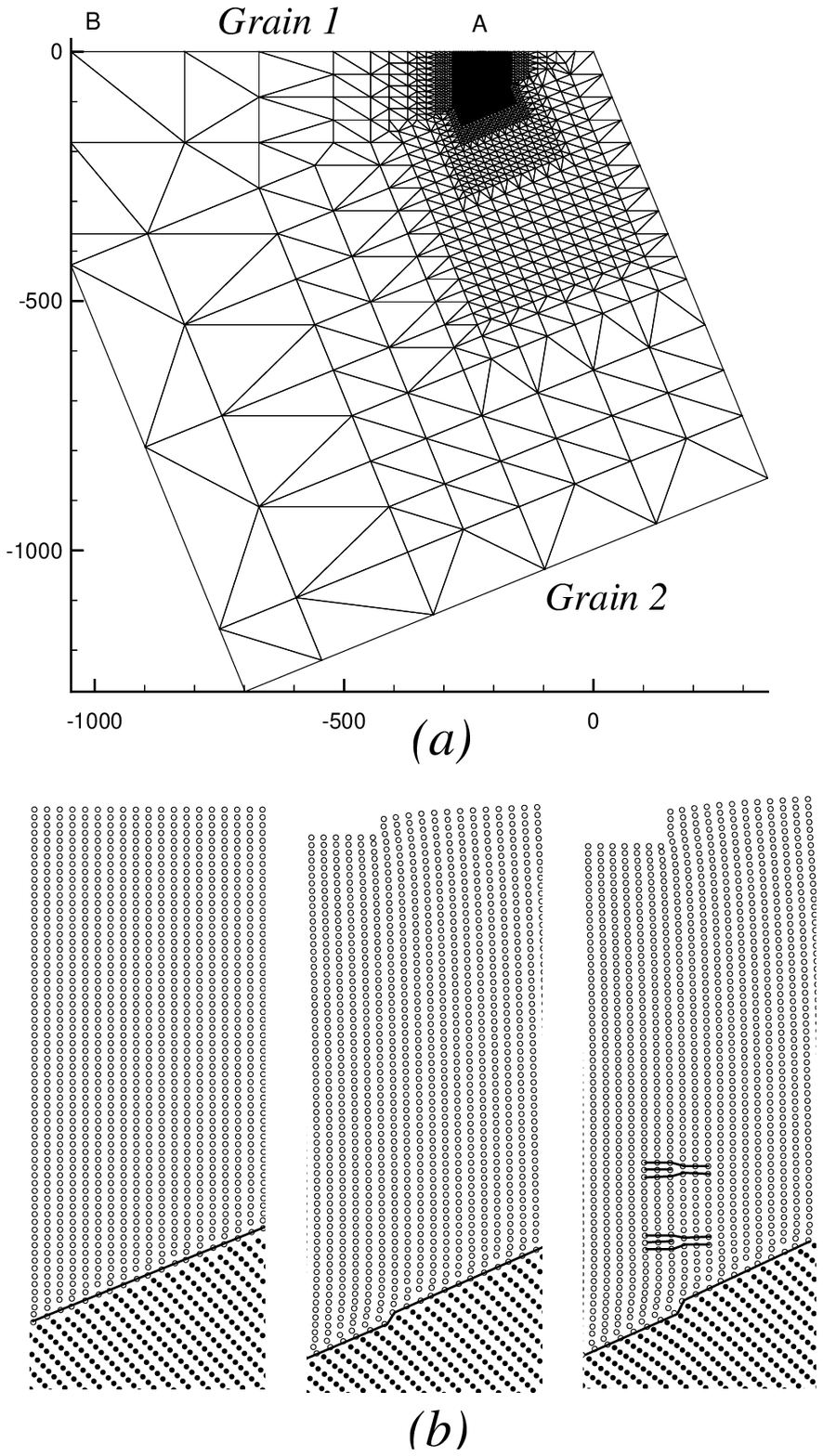}
\caption{Example of a multiscale simulation using the
quasicontinuum method. (a) Finite-element mesh used to model
dislocation-GB interaction. The surface marked AB is rigidly
indented in order to generate dislocations at A (distance in \AA).
(b) Snapshots of atomic positions at different stages in the
deformation history. Absorption of the first pair of dislocations
at the GB results in a step, while the second pair form a pileup.}
\label{GB}
\end{figure*}

\begin{figure*}
\includegraphics[width=400pt,angle=270]{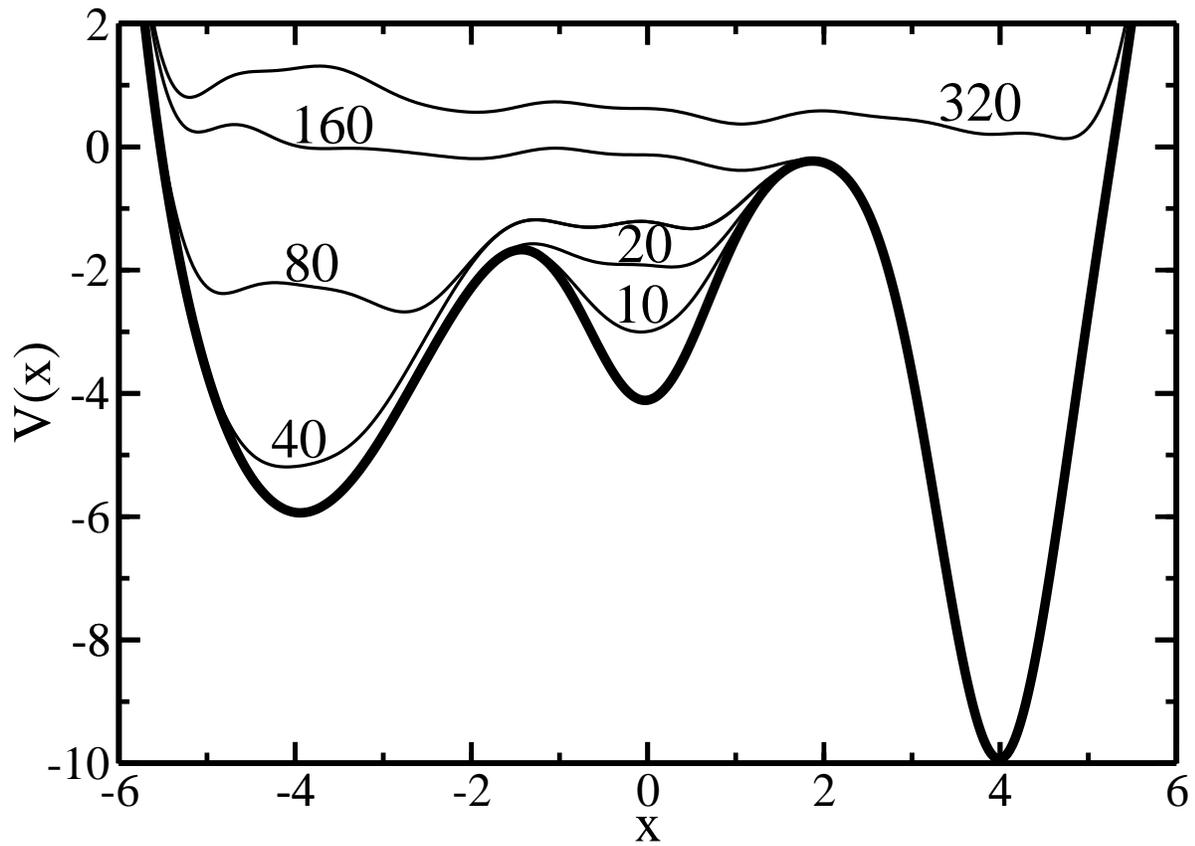}
\caption{ Time evolution of the sum of a one-dimensional model
potential V($\sigma$) and the accumulating Gaussian terms of Eq.
(\ref{gauss}). The dynamic evolution (thin lines) is labelled by
the number of dynamical iterations in Eq. (\ref{dyn}). The
starting potential (thick line) has three minima and the dynamics
is initiated in the second minimum.} 
\label{gauss1}
\end{figure*}

\end{document}